\documentclass[preprint]{aastex62}
\usepackage{graphicx}
\usepackage{multirow}
\usepackage{longtable}
\usepackage{epstopdf}
\usepackage{array}
\shorttitle{On LAT GRB afterglow spectra}
\shortauthors{He \etal}

\begin{document}

\def\apj{{ApJ}}
\def\mnras{{MNRAS}}
\def\aa{{A\&A}}
\def\Nature{{Nature}}
\def\GCN{{GCN Circ}}
\def\PRD{{Phys. Rev. D}}
\def\PRL{{Phys. Rev. Lett}}
\def\etal{{\it et al.}}
\def\gr{{$\gamma$-ray}}

\title{Is an upturn commonly seen in Fermi-LAT GRB afterglow spectra?}

\author{He, Xin-Bo, Tam, Pak-Hin Thomas, Long, Guang-Bo, and Zhang, Yong}
\email{Email: tanbxuan@sysu.edu.cn (P.H.T. Tam)}

\affiliation{Sun Yat-sen University}

\begin{abstract}
We analyzed 199 GRBs detected by Fermi-LAT during the years 2008--2019. We found 67 photons at energies $\geq$10~GeV, which come from 34 GRBs. Out of these 34 GRBs, Fermi-LAT detects significant (\emph{TS}$\geq$4) afterglow 0.1--200~GeV photons from 25~GRBs. We present time-integrated 0.1--200~GeV spectra of these 25 GRBs. The spectra of a significant fraction (9/25) of these GRBs revealed a harder component above an energy break of 0.3-2 GeV. While shock synchrotron may account for the photons at the lower energy end, high energy photons above the break is naturally explained by synchrotron self-Compton (SSC) emission. We perform broadband model fit to the X-ray-LAT emission of GRB~131231A. Comparing the afterglow spectra of these 25 GRBs with other Fermi-LAT detected GRBs, we found that the power-law index distribution is similar for the two populations. This may indicate that the additional high-energy component may also exist in Fermi-LAT GRBs in general. 
%Generally, the afterglow of GRBs is described by synchrotron emission,
%After MAGIC and HESS observed VHE photons in afterglow phase, it is well accepted that SSC emission are detected in the afterglow phase, at least for some GRBs.

\end{abstract}


\section{Introduction}
\label{sect1}
Gamma-ray bursts (GRBs) consist of short (tens of milliseconds to thousands of seconds) and bright prompt emission at $\sim$keV to MeV energies~\citep{Ree92,Meszaros97}, followed by longer (days to months) afterglow phase at lower energies~\citep{Costa97}. Generally, the prompt emission is expected to be produced by some internal dissipation mechanisms such as collisionless shock in jet~\citep{Ree94}, and the afterglow emission, from radio up to X-rays, can be described by synchrotron emission of external shock with surrounding medium~\citep{Meszaros97}.

%Fermi-LAT have detected GRBs with high energy (HE, $\geq$ 100 MeV) and even several very high energy (VHE, $\geq$ 10 GeV) photons~\citep{Meegan1998,lat_grb_cat2}.

 {\it Fermi Gamma-ray space Telescope\footnote{\url{https://fermi.gsfc.nasa.gov}}}~\citep{Fermi}, launched in 2008, plays an important role in the detection, as well as the spectral and timing characterisation of GRBs, with its two onboard instruments: the Large Area Telescope (LAT, 30 MeV -- 500 GeV)~\citep{Fermi-LAT} and Gamma-ray  Burst  Monitor (GBM, 8 keV -- 40 MeV)~\citep{Fermi-GBM}. At $>$30~MeV, the Fermi-LAT can monitor GRBs in both the prompt and afterglow phases. It has detected 186 GRBs~\citep{lat_grb_cat2} in the first decade, or 8\% of all GRBs at $>$100~MeV energies~\citep[see, e.g.][]{lat_grb_cat,lat_grb_cat2}. $\ga$10 GeV photons have been detected from GRBs 080916C~\citep{Abdo2009}, 090902B (and 090328, 100414, 110721, 110731, 130427, 140619B)~\citep{Panaitescu2017}, 090926A~\citep{Yassine2017}, 130427A~\citep{Tam2013, 130427A}, 130907A~\citep{Tang2014}, 131231A~\citep{Liu2014}, 160509A~\citep{Tam2017}. %, 180720B~\citep{HESS} and 190114C~\citep{MagicB,Fraija2019,Derishev2019,Wang2019,Zhang2019}.

Furthermore, Cherenkov telescope arrays MAGIC and HESS have recently detected very high-energy (VHE; $\ga$100~GeV) emission from several GRBs, largely at the afterglow phase: GRB~180720B~\citep{HESS}, GRB~190114C~\citep{MagicA}, and GRB~190829~\citep{GCN25566}). In the case of GRB~190114C, the highest observed photon energy reach $\sim$1~TeV~\citep{MagicA}.

Concerning the origin of the afterglow GeV emission, synchrotron emission from external shock electrons becomes the ‘standard’ radiation mechanism in the Fermi era but there exists a maximum synchrotron energy, typically $\la$10 GeV~\citep[e.g.,][]{KBD10, PiranNakar10}. It is a great challenge for the traditional synchrotron mechanism to explain the $\ga$10 GeV emission, since the flux emitted by this mechanism above several GeV (the maximal cutoff energy) should be very small. The synchrotron self-Compton (SSC) component is expected to dominate in the GeV-TeV energy range~\citep{Meszaros2000,Sari2001,Zhang2001}, implying that the spectra of such GRBs should have two peaks in the broad-band SED: one at the X-ray band, and another peak at the TeV \gr~band.
 
\citet{Tam2013}, for the first time, found an extra hard component in the late LAT afterglow in GRB~130427A. Following this, \citet{Panaitescu2017} have found a sample of LAT GRBs whose afterglow spectrum is better described by broken power-law (BPL) with a harder high-energy photon index ($\beta$) than low-energy photon index ($\alpha$), providing a strong evidence of hardening spectrum above a few GeV, indicating that an inverse Compton spectral component may exist in GRB afterglow phase. The analysis by \citet{Panaitescu2017} is aperture photometry-based, and is focused on the first 1000s after the bursts. 

Given the detection of $\geq$100~GeV photons by Cherenkov telescope arrays from GRB afterglows up to a day, it is desirable to look for photons with the highest energies accessible to Fermi-LAT, e.g., 10--200~GeV, up to one day post-burst. Indeed, Fermi-LAT covers the spectral range between the two peaks (synchrotron and inverse-Compton) expected in the afterglow spectra, and has likely been seen in GRB~190114C~\citep{MagicB}. In this work, we expand the above works to all LAT-detected GRBs, and up to one day after the burst, using maximum likelihood analysis.
 
\section{Datasets and methods}
\label{sect2}

\subsection{The Sample and the Search of $\ge$ 10 GeV photons}
\label{sect2.1}
From 2008 to 2020, Fermi-LAT has detected more than 200 GRBs. The information about the Fermi-LAT GRBs can be obtained from the second Fermi-LAT GRB catalog~\citep{lat_grb_cat} and the Fermi website\footnote{\url{https://fermi.gsfc.nasa.gov/ssc/observations/types/grbs/lat_grbs}}. In June 2019, Fermi-LAT team released the Second Fermi-LAT GRB Catalog (the first decade of GRBs detected by the Fermi-LAT), containing 186 LAT GRBs, from a search of a total of 2357 GRBs~\citep{lat_grb_cat2}. In order to enlarge the sample size of possible GRBs with hard photons as much as possible, we also consider GRBs listed in the webpage of Fermi LAT GRBs\footnote{\url{https://fermi.gsfc.nasa.gov/ssc/observations/types/grbs/lat_grbs/table.php}, retrieved on 2019 March 31}. It contains 146 GRBs (including some GRBs in the second Fermi-LAT GRB Catalog). GRB~180720B and GRB~190114C are the first two GRBs identified to emit $\sim$TeV emission, so they are also included here. In the end, the GRB sample considered in this work contains a total of 199 GRBs.

\begin{figure}
\centering
\includegraphics[scale=0.4]{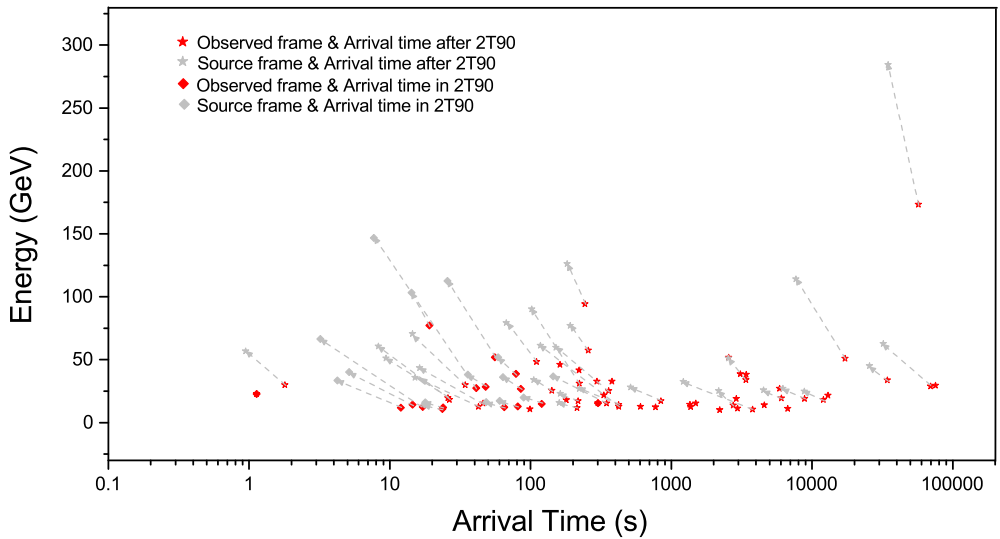}
\caption{All the GeV photons we got from Fermi-LAT GRBs during the time interval from onset to 1day. We divide the photons arrived time by 2T90. Red and Grey star symbols represent the photons arrived within and after 2T90, respectively. The photons energy and arrival time in source frame are calculated if the redshifts are given. The diamond symbols are the photons in source frame. The arrowheads link photons in observed frame to those in source frame.
}
\label{fig1}
\end{figure}
Fermi-LAT data were downloaded from the {\it Fermi-LAT database\footnote{\url{https://fermi.gsfc.nasa.gov/ssc/data/access/}}}. Fermi Science Tools version 1.2.1 was used to reduce and analyze the data in 0.1-200 GeV. For the event class, we selected "P8R3\_SOURCE" for the GRB spectra analysis, as we focus on longer than 200~s intervals which are typical for afterglows. The IRF used in the analysis is ``P8R3\_SOURCE\_V2\_v1". The zenith angle is constrained to be smaller than 100 deg to exclude the gamma-ray emission from the Earth.

We performed unbinned maximum-likelihood analyses (gtlike) to characterize spectra in the time range from GRB onset to 1 day thereafter. The reasons to derive time-integrated spectra instead of time-resolved ones are two-fold: (1) the lack of photon statistics for many of the LAT GRBs especially in the afterglow phase; (2) there is no strong indications for change of spectral index during the afterglow phase~\citep{lat_grb_cat}. The selected ROI is 10 deg centered on the position of target GRBs. To suppress the background, the newest diffuse model gll\_iem\_v07.fits (Galactic diffuse emission) and iso\_P8R3\_SOURCE\_V2\_v1.txt (isotropic diffuse component) are used in our analysis, sources in the Fermi catalog (4FGL) are included as background sources. Generally, the LAT emission in afterglow phase can be fit well with a power-law (PL) model:
\begin{equation}
\frac{dN}{dE} = N_0 \left(\frac{E}{E_0}\right)^{-\Gamma},
\end{equation}
Here we allow the spectral parameters of galactic, isotropic component, and those of the GRBs to vary and fix other sources' parameters. The best fitting is determined by maximising the test-statistics (\emph{TS}) value, which is defined as:
\begin{equation}
TS = 2~log \frac{L_\mathrm{null}}{L_\mathrm{source}},
\end{equation}

Next, we used gtsrcprob {\it (which computes the probability of source association in the model for each event)\footnote{\url{https://raw.githubusercontent.com/fermi-lat/fermitools-fhelp/master/gtsrcprob.txt}}} tool to estimate the probability of each photon coming from a particular source, $P_\mathrm{source}$. We check all the 199 GRBs to look for $\geq$10~GeV photons associated with a GRB, which we define as those with $P_\mathrm{GRB}\geq97\%$. Finally, we obtain 67 photons above 10 GeV from 34 GRBs, which are listed in Table~1 and Fig.~1. \emph{These 34 GRBs are potential GRBs equipped with very high-energy photons.} Our result is consistent with Fermi-LAT team's analysis~\citep{lat_grb_cat2}. In the 34 GRBs, only one GRB (GRB~090510A) is short-duration, while others are long GRBs. Three categories can be seen:
\begin{itemize}
\item for GRB~160625B, 140619B, 110903A, 080916C, their $\geq$10~GeV photons arrived in the prompt phase;
\item for most GRBs, the $\geq$10~GeV photons arrived after the prompt phase;
\item for 160509A, 131231A, 090902B, their $\geq$10~GeV photons arrived in both prompt and afterglow phases.
\end{itemize}
All photons have energy below 100 GeV (in the observer's frame), except for an interesting event -- a 173 GeV photon arrived at 57071~s after trigger with about 97$\%$ probability being associated with GRB~131231A. The photons' energy, arrive time and duration in observed frame and source frame are displayed in Figure~\ref{fig1}. It is hard to explain the late time VHE photons by synchrotron emission from external shock, as the maximum (cutoff) photon energy radiated by synchrotron is $\sim$ 5 GeV~\citep{PiranNakar10,Abdo2009,Duran2011}. It is therefore desirable to see if an additional high-energy spectral component can exist in the Fermi-LAT spectra, especially at the high energy end above 10~GeV. In the following section, we turn to the analysis of Fermi-LAT spectra in the afterglow era.

 \begin{figure}
  \centering
\includegraphics[width=120mm,height=80mm]{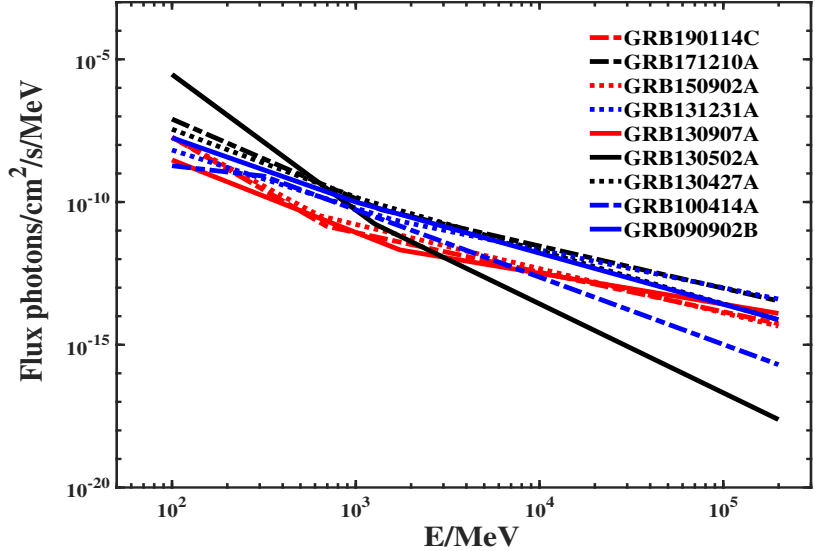}
\caption{the best-fitting photon spectra of afterglows with $\Delta$\emph{TS} $\ge$ 3 in the BPL model.}
\label{combined}
\end{figure}

\subsection{Spectral analysis of GeV GRB afterglows with $\geq$10~GeV photons}
\label{sect2.3}
To obtain the afterglow spectra, the time interval 2T90-86.4~ks is adopted to exclude the prompt emission phase of GRBs. Out of the 34 GRBs with $\geq$10~GeV photons, only 25 of them were detected in the time interval after 2$\times$T90 (the 25 ``GeV afterglow GRBs detected at $\geq$10~GeV") with \emph{TS} $\ge$4. We then use the PL model to fit the observed spectra of the GeV-GRB afterglows, and present them in Figures~\ref{sample1} and~\ref{sample2}. Following \citet{Tam2013} and \citet{Panaitescu2017}, we also test the broken power law (BPL) model:
 \begin{equation}
\frac{dN}{dE} = N_\mathrm{0} \left\{
\begin{array}{ll}
	\left(\frac{E}{E_\mathrm{b}}\right)^{-\alpha}\,\mathrm{if}\,E<E_\mathrm{b} \, , \\
	\left(\frac{E}{E_\mathrm{b}}\right)^{-\beta}\,\mathrm{if}\,E\geq E_\mathrm{b} \, ,
\end{array}
\right.
\end{equation}
where $E_\mathrm{b} $ is the break energy. The fitting results are shown in Figures~\ref{sample1} and \ref{sample2}.

In order to compare the two models: PL versus BPL, quantitatively, we define $\Delta TS = TS_{BPL}-TS_{PL}$. When $\Delta TS>3$, the BPL model is preferred, and vice versa. We find that 10 of the spectra can be fitted better with BPL model than with PL model, and for the remaining 15, both models give comparable goodness-of-fit (and thus PL is preferred given its simplicity). For the 10 GRBs, there is a spectral break within the energy range 0.1--200~GeV. Among these spectra, the spectra show hardening above a break energy in 9 of them, i.e., the high-energy photon index ($\beta$) is harder than the low-energy photon index ($\alpha$), and only in GRB~100414A the spectrum becomes softer at high energies. We plot the spectra with $\Delta$ \emph{TS} $\ge$ 3 in Figure~\ref{combined}. One can see that the break energy $E_\mathrm{b}$ is around 1 GeV for most of these spectra. The upturn of afterglow spectra is most significant (with the highest $\Delta$\emph{TS} values) in GRB~131231A, GRB~150902A and GRB~190114C.

\begin{figure}
 \centering
\includegraphics[width=120mm,height=80mm]{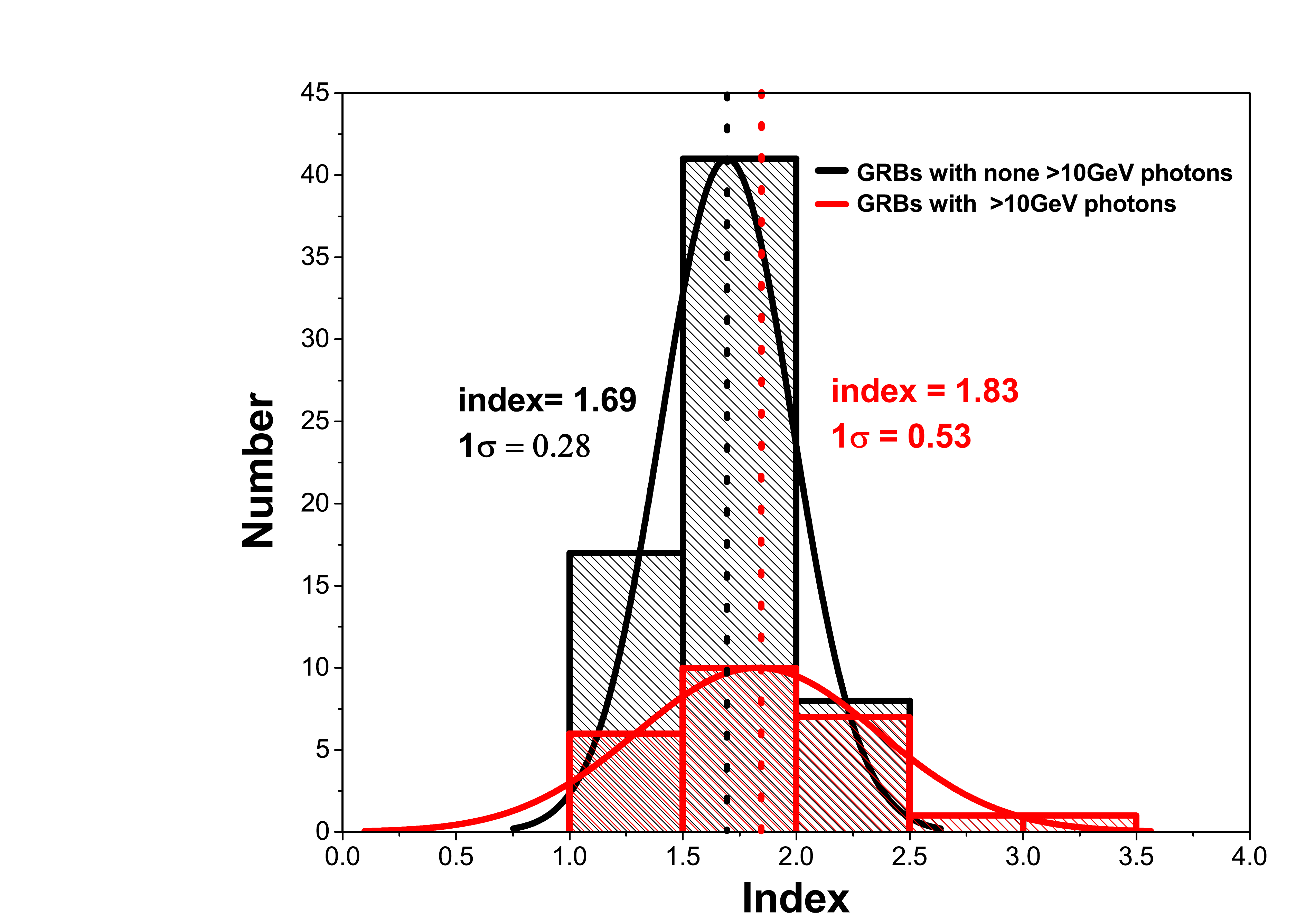}
\caption{the distribution of spectral index from 174 typical GRB afterglows (black line) and 25 GeV GRBs (red line) respectively. The spectra of afterglows from 2$T90$ to 86.4ks are fitted with the PL model in 0.1-200 GeV.}
\label{distribution}
\end{figure}

 To compare the spectral indices between the 25 ``GeV afterglow GRBs detected at $\geq$10~GeV" with other LAT GRBs, we plot the index distribution in Figure~\ref{distribution}. The average spectral index of the 25 ``GeV afterglow GRBs" ($\Gamma=1.83\pm0.53$) is comparable or even slightly softer than that of the typical Fermi-LAT GRBs ($\Gamma=1.69\pm0.28$). If the GeV afterglow spectra only result from the synchrotron emission, it should be exponential cutoff above 10 GeV and the average spectral index of GeV GRBs should be softer (e.g., $\Gamma\ga$2) than the observed values. This may indicate the existence of $\geq$10~GeV photons or even a hard component at high energies for a typical LAT GRB afterglow.
 
\section{Discussion}
%\subsection{Interpreting the spectra upturn with an SSC component origin}

The upturn of afterglow spectra in some GRBs, especially for GRB~131231A, GRB~150902A and GRB~190114C (the three cases having the highest $\Delta$\emph{TS} values), indicate that an extra spectral component should exist at VHE. The most natural candidate for this component is the SSC radiation in the external forward shock~\citep[see][in the case of GRB~190114C]{MagicB}. The SSC emission is generally used to explain the VHE $\gamma$-ray observation in the high energy astrophysical phenomenon. In the GRB afterglows, it has been predicted over the past two decades~\citep{Panaitescu1998,Wei1998,Sari2001,Zhang2001,Meszaros2004,
Fan2008,Galli2008,Nakar2009,PiranNakar10,Lemoine2015} and the VHE emission detected in the GRB~180720B and GRB~190114C provides evidence for an SSC component origin~\citep{MagicB,HESS}.

In the syn$+$SSC scenario, the spectrum below $E_\mathrm{b} $, the synchrotron photon index should be larger than 2, the SSC emission kicks in to make the spectra ``anomalously" turn up above a few GeV. This is broadly consistent with the 9 GRBs with an upturn at $\sim$1~GeV. %However, for GRB~100414A, the break may be caused by the cutoff of synchrotron emission at high energy.

In order to investigate whether the SSC emission can explain the new component, we fit the observations of GRB~131231A whose spectrum upturn is most significant in terms of $\Delta TS$ with a simple SSC model presented by Sari \& Esin~\citep{Sari1998,Sari2001}. This model gives an analytic approximation of synchrotron and SSC spectra ignoring the KN effects. This approximation is adequate given the limited photon statistics up to 200~GeV.

\begin{figure}[!h]
\centering
%  \begin{minipage}[b]{0.5\textwidth}
%    \centering
%%    \includegraphics[width=8cm,height=3.0cm]{swift1312.eps}
%\includegraphics[angle=-90, scale=0.3]{swift1312.eps}
%  \end{minipage}\\
%  \begin{minipage}[!h]{0.427\textwidth}
%    \centering
%%    \includegraphics[width=7.93cm,height=3.7cm]{fermi1312.eps}
%\includegraphics[angle=-90, scale=0.3]{fermi1312_2.eps}
%  \end{minipage}
\includegraphics[angle=-90, scale=0.45]{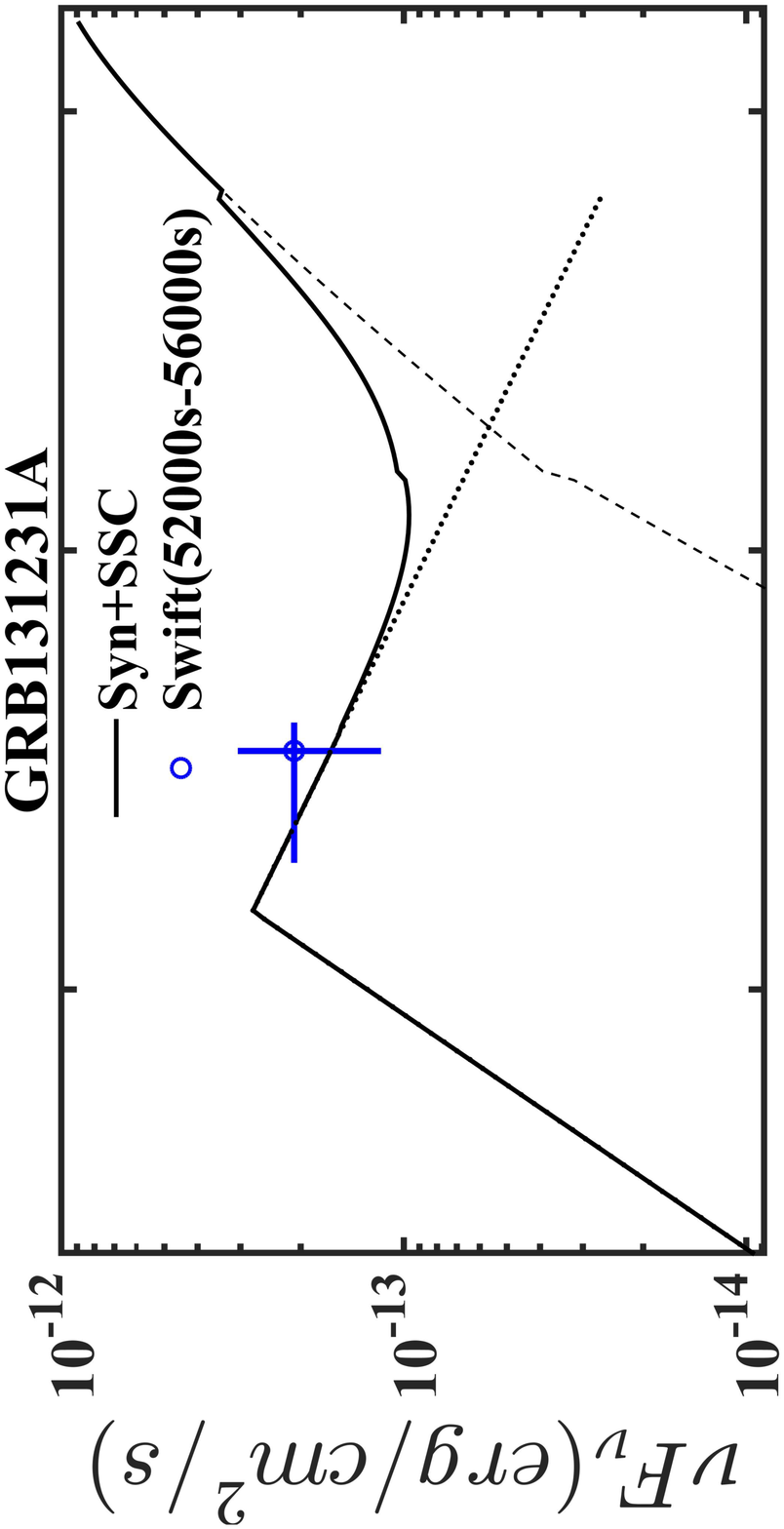}
\includegraphics[angle=-90, scale=0.445]{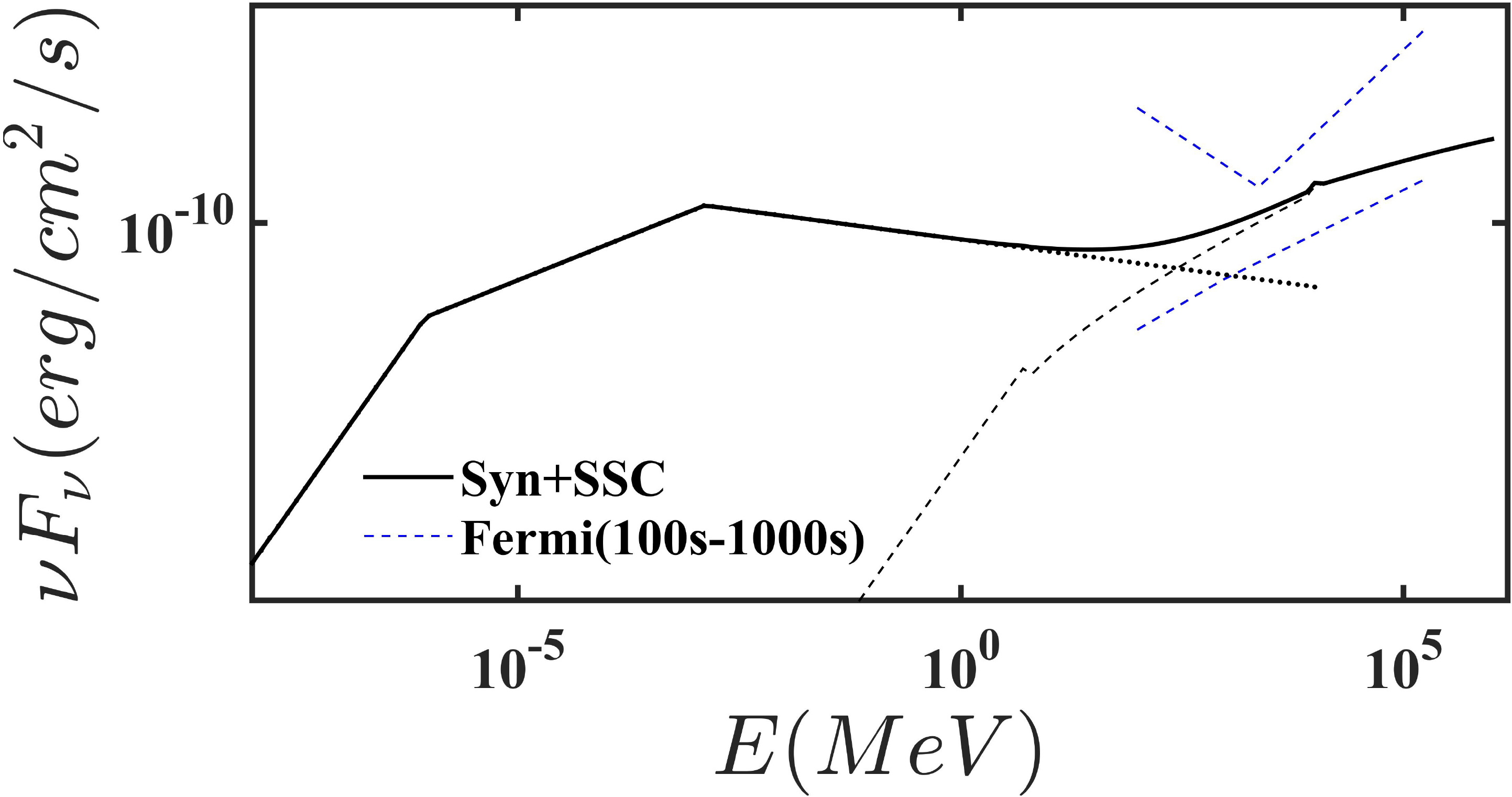}
  \caption{\small{Modelling the broadband spectra of GRB~131231A in the time intervals 100-1000s and 52000-56000s. Thick black curves represent the theoretical spectra of synchrotron plus SSC corresponding to slow cooling in the external-forward shocks scenario. Dotted line and dashed line is synchrotron and SSC components, respectively. The adopted parameters: $\epsilon_e$=0.1, $\epsilon_B$=0.0015, p=2.25, n=1 $cm^{-3}$, $\eta$=0.1}. The meaning for the observed data is labeled in the picture.}
  \label{modeling}
\end{figure}

The result of modeling to the Swift and Fermi data of GRB~131231A, within the framework of the theory of afterglow emission from external forward shocks~\citep{Sari1998,Sari2001}, is shown in Figure~\ref{modeling}. We find that the best fitting model parameters are in the range of typical values for GRB afterglows and acceptable electron spectra that can model the upturn of Fermi observed spectra are obtained. The relativistic shock propagates into a constant surrounding density n$\thickapprox$1 $cm^{-3}$, accelerating the electrons and forming a power law distribution of Lorentz factor $\gamma_e$ with index p$\thickapprox$2.25. A constant fraction $\epsilon_e\approx0.1$ of the shock energy goes into the electrons and about 10\% of the electron energy is radiated away via Syn+SSC emission. A fraction $\epsilon_B\approx0.0015$ of the shock energy go into amplifying the magnetic fields behind the shock. Corresponding to these values of fitting parameters, observed time and flux (1.52$\times 10^{-4}$ erg $cm^{-2}$), the emitting relativistic electrons are in the regime of slow cooling. We note that $\epsilon_e\gg\epsilon_B$, thus the necessary condition for efficient production of SSC radiation can be satisfied~\citep{Sari1998,Sari2001}. These results are consistent with a previous work by~\citet{Liu2014}, especially on the origin of the LAT afterglows before 1000s post-burst.

Based on the spectra shown in Figure~\ref{combined}, the spectra at high energies are harder than low energies, if both components are originated from the same shock, the low energy component is synchrotron emission and higher energy components are SSC emission. In this case, the synchrotron emission should be produced above the cooling break, and $\alpha = p/2$. and spectra of SSC in high energy component should be $\beta= (p-1)/2$, and $\alpha - \beta $ should be $1/2$~\citep{Panaitescu2017}. In our BPL result, we can find that GRB~160310A ($\sim 0.7$), GRB~130427A ($\sim0.52$) and GRB~090902B ($\sim 0.46$) can be satisfied with $\alpha - \beta = 0.5$.

We infer the high energy emission of GRB~100414A and GRB~101014A are produced by synchrotron emission, given their soft spectra. The photon index distribution of the 25 ``GeV afterglow GRBs detected at $\geq$10~GeV"  is similar to other 174 Fermi-LAT GRBs, and both are indeed rather hard ($\Gamma\,<$2) at the LAT band. This result lends support to the idea that these 25 GRBs do not form a distinctive class of their own, and their LAT afterglow spectra are not dissimilar to most other Fermi-LAT GRBs. This prompts us to speculate on the possibility that LAT GRB afterglows are hard in general (in the 0.1--200~GeV band, allowing for the existence of an SSC component), and therefore may emit photons at $\geq$10~GeV energies.

If this is true, then the reason of non-detection of the very high-energy ($\geq$100~GeV) photons of most GRBs, may be due to Klein-Nishina cut-off~\citep{Nakar2009,aliu14} , internal $\gamma$$\gamma$ absorption\citep{Panaitescu2017,Derishev2019}, or attenuation by the extragalactic background light~\citep{MagicA}. This can be ultimately tested by the LHAASO-WCDA detector~\citep{lhaaso} and the CTA array~\citep{CTA}.

\section{Conclusions}
In this work, we perform an analysis of all Fermi-LAT GRBs. While \citet{Panaitescu2017} have focused on the first 1000~seconds, we reanalysis all GRB afterglows detected by the Fermi-LAT in the energy band of 0.1--200 GeV, for a time frame up to 1 day (i.e., 86.4~ks) after the burst onset. 

We found 67 photons at energies $\geq$10~GeV, which come from 34 GRBs. Out of these 34 GRBs, Fermi-LAT detects significant (\emph{TS}$\geq$4) afterglow 0.1--200~GeV photons from 25~GRBs. In this work, we perform time-integrated 0.1--200~GeV spectra of these 25 GRBs, finding that the spectra of the 25 afterglows are rather hard (average spectral index is $\sim$1.8). 10 of the 25 GeV GRBs are characterised by a BPL spectrum with a 0.3--2~GeV break, and 9 of them turn up. The upturn of GRB~131231A can be explained by the SSC model under typical parameter values of GRBs. Our results suggest that the spectra upturn presented in Fermi-LAT data is likely to be resulted from the SSC emission and the SSC emission could be commonly produced in GRBs.

\section*{Acknowledgments}
This work is supported by the National Natural Science Foundation of China (NSFC) grants 11633007, 11661161010, and U1731136.
This work made use of the LAT data and science tools available at the Fermi Science Support Center, https://fermi.gsfc.nasa.gov/ssc/data/.

\label{sect3}

\begin{figure*}
\includegraphics[width=60mm,height=48mm]{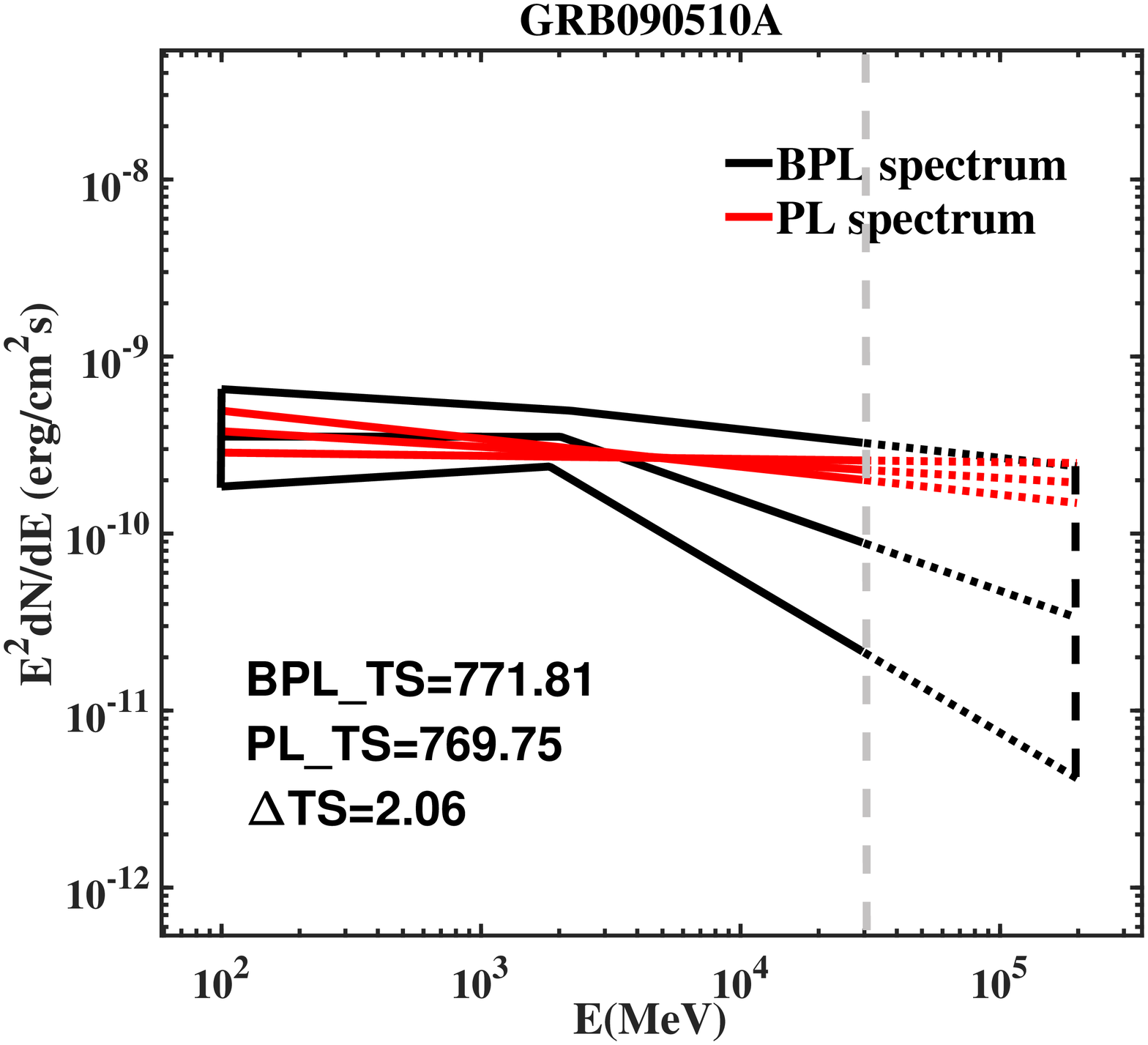}
\includegraphics[width=60mm,height=48mm]{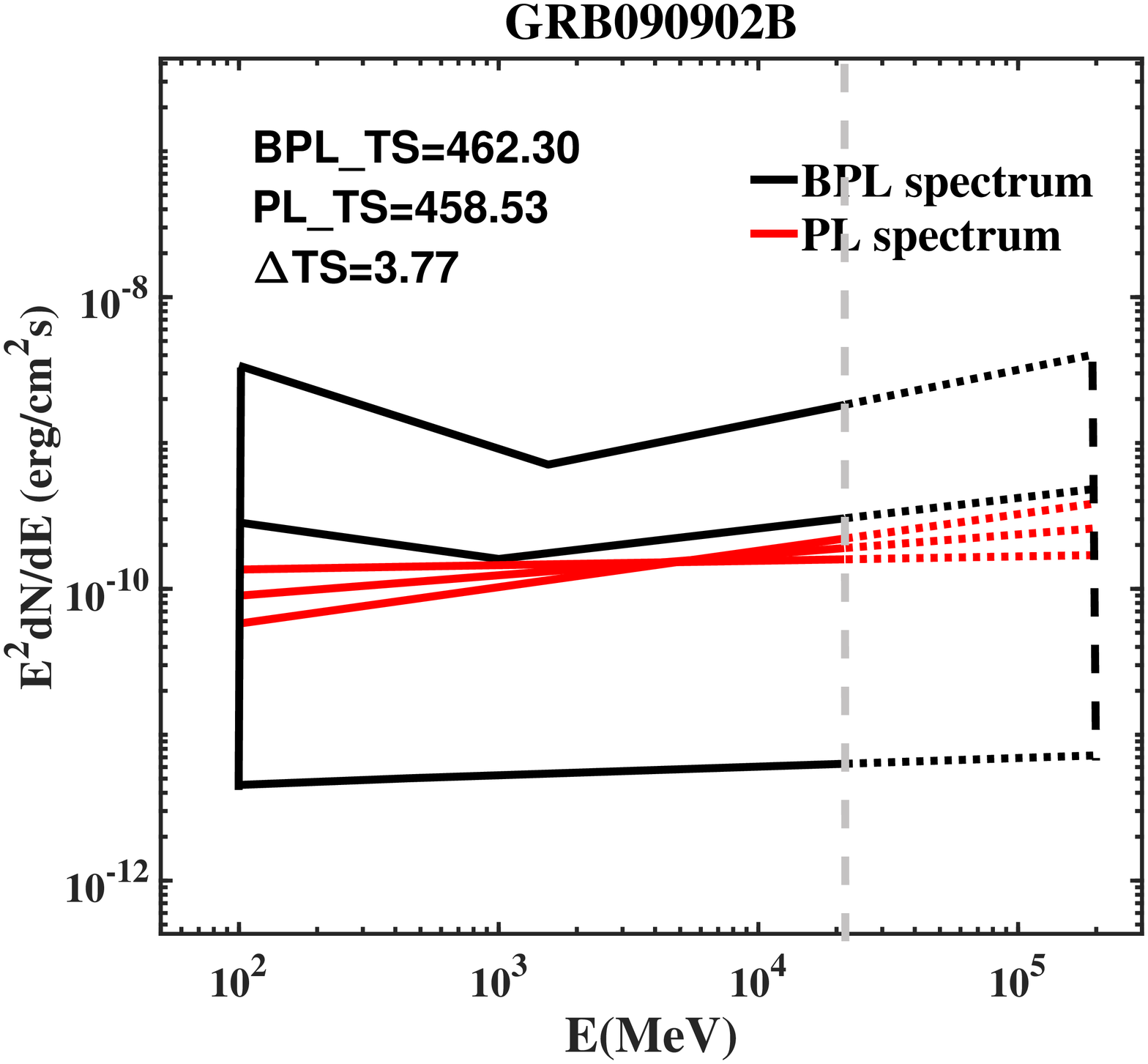}
\includegraphics[width=60mm,height=48mm]{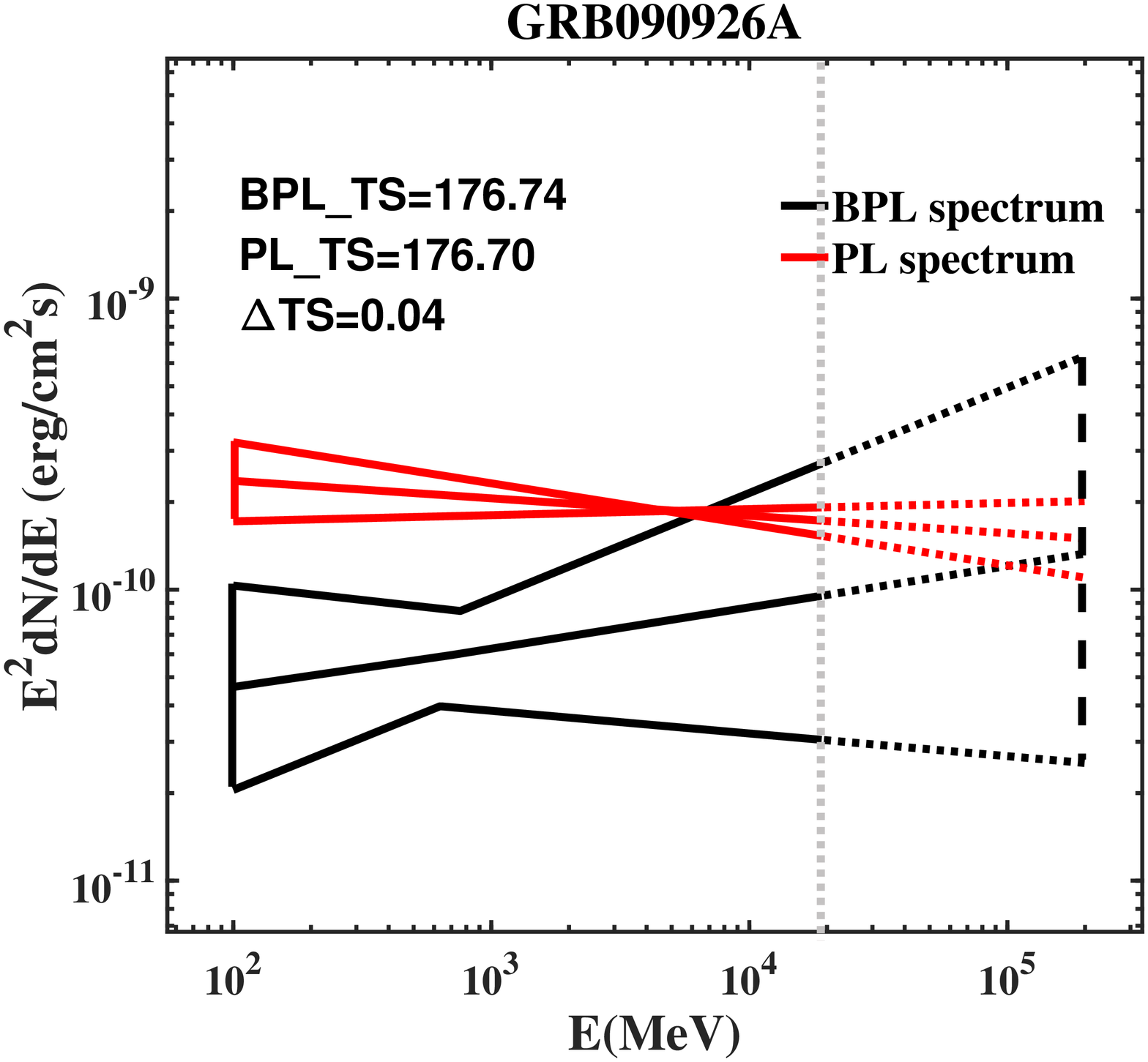}
\includegraphics[width=60mm,height=48mm]{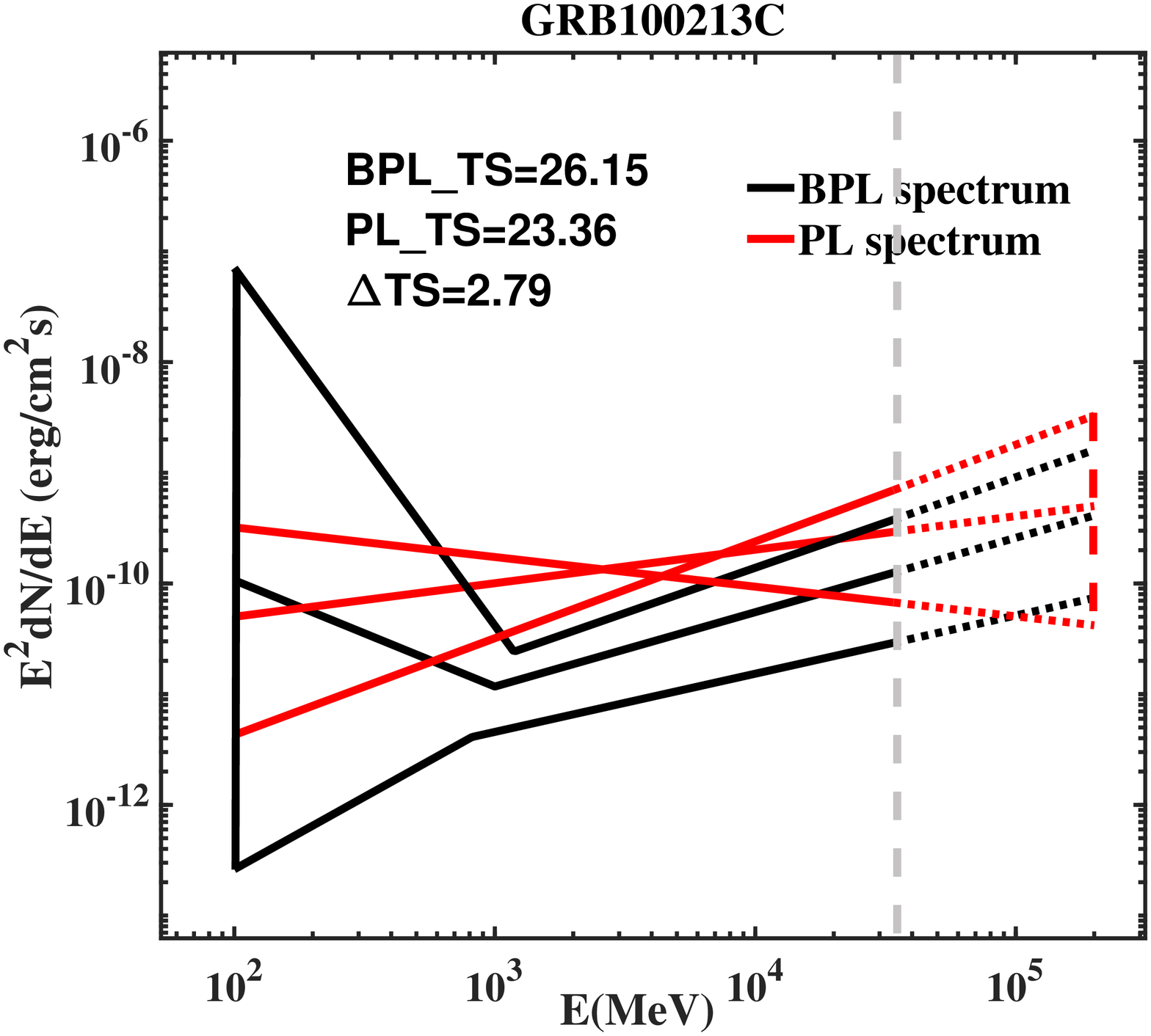}
\includegraphics[width=60mm,height=48mm]{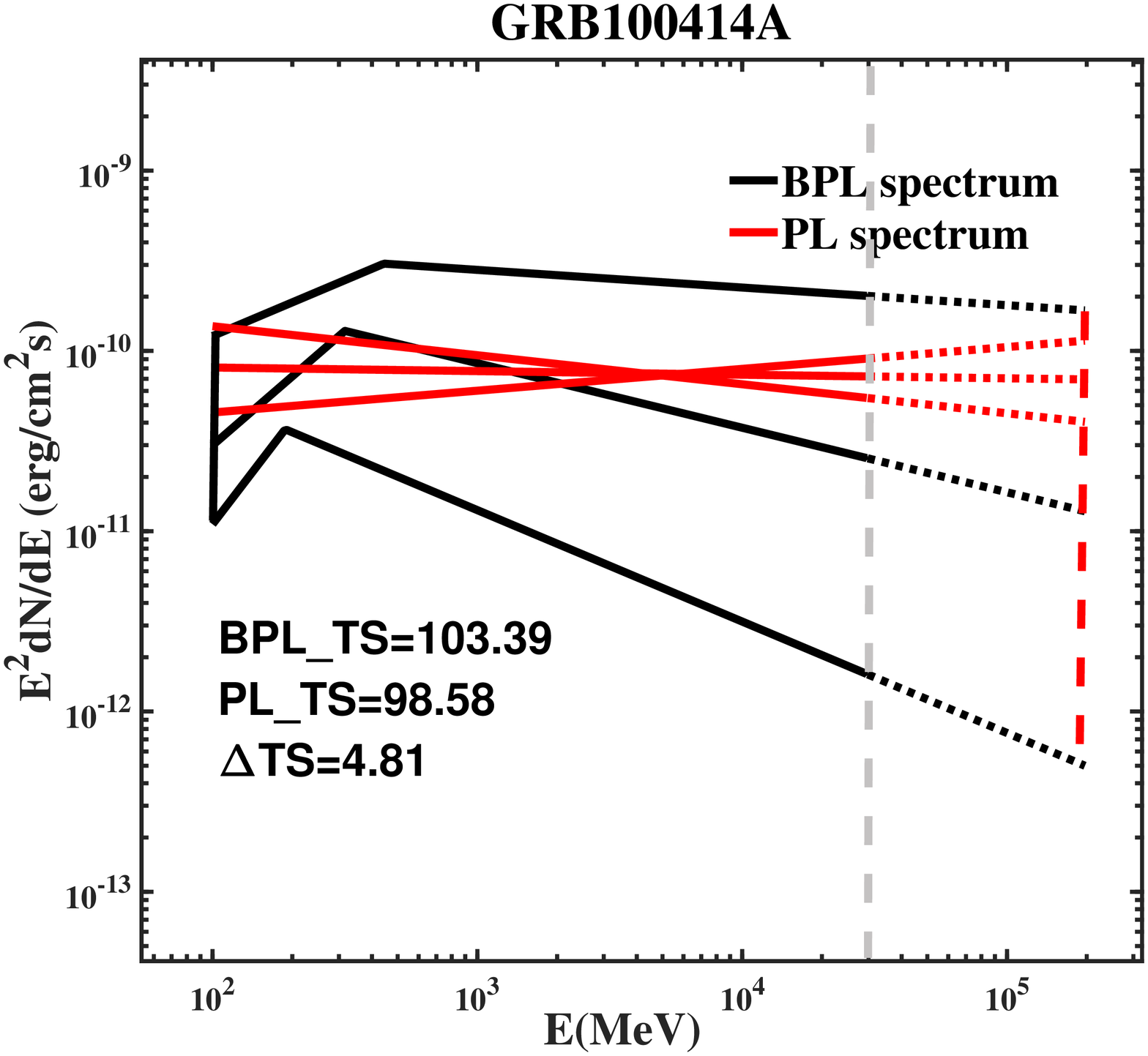}
\includegraphics[width=60mm,height=48mm]{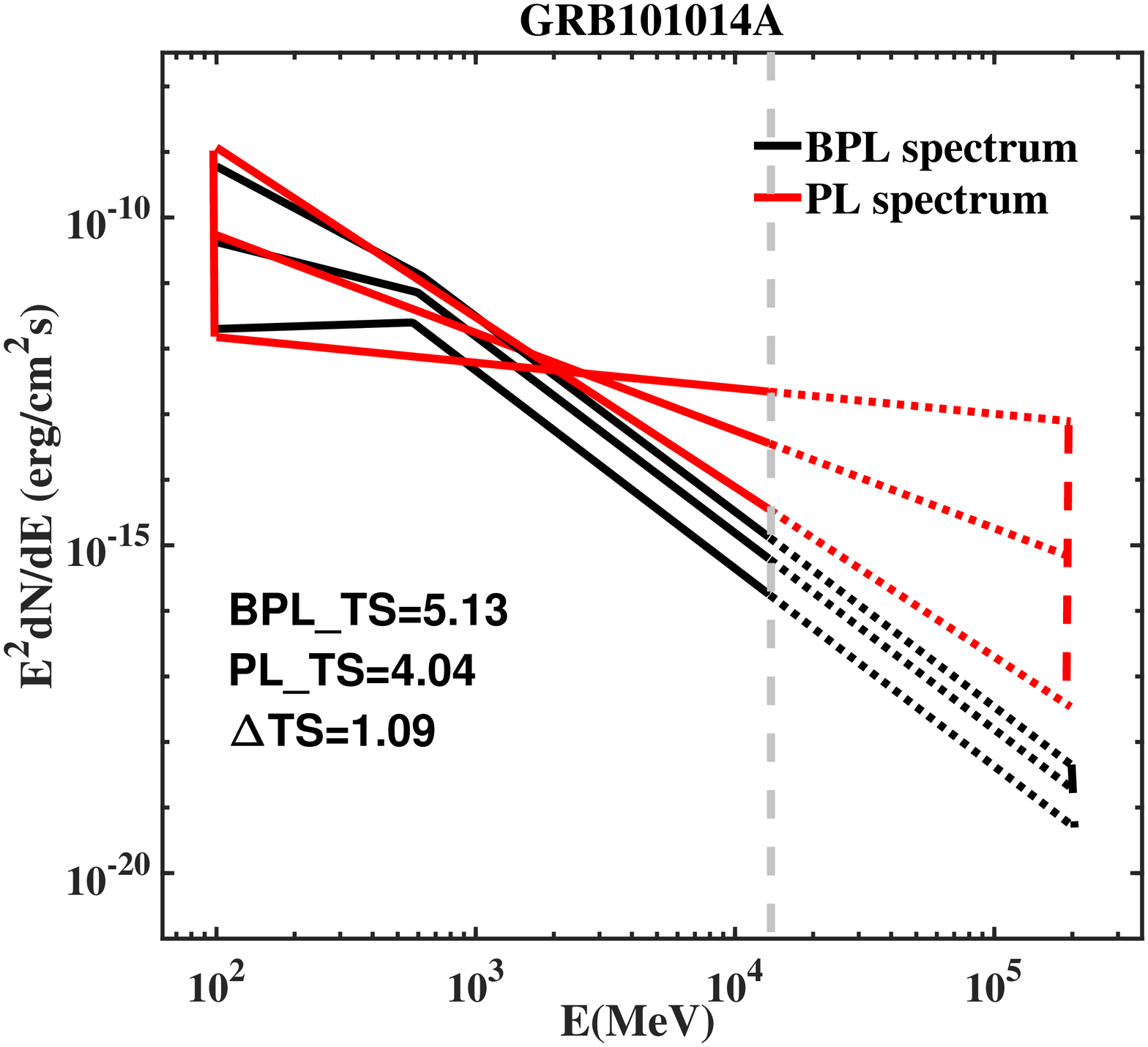}
\includegraphics[width=60mm,height=48mm]{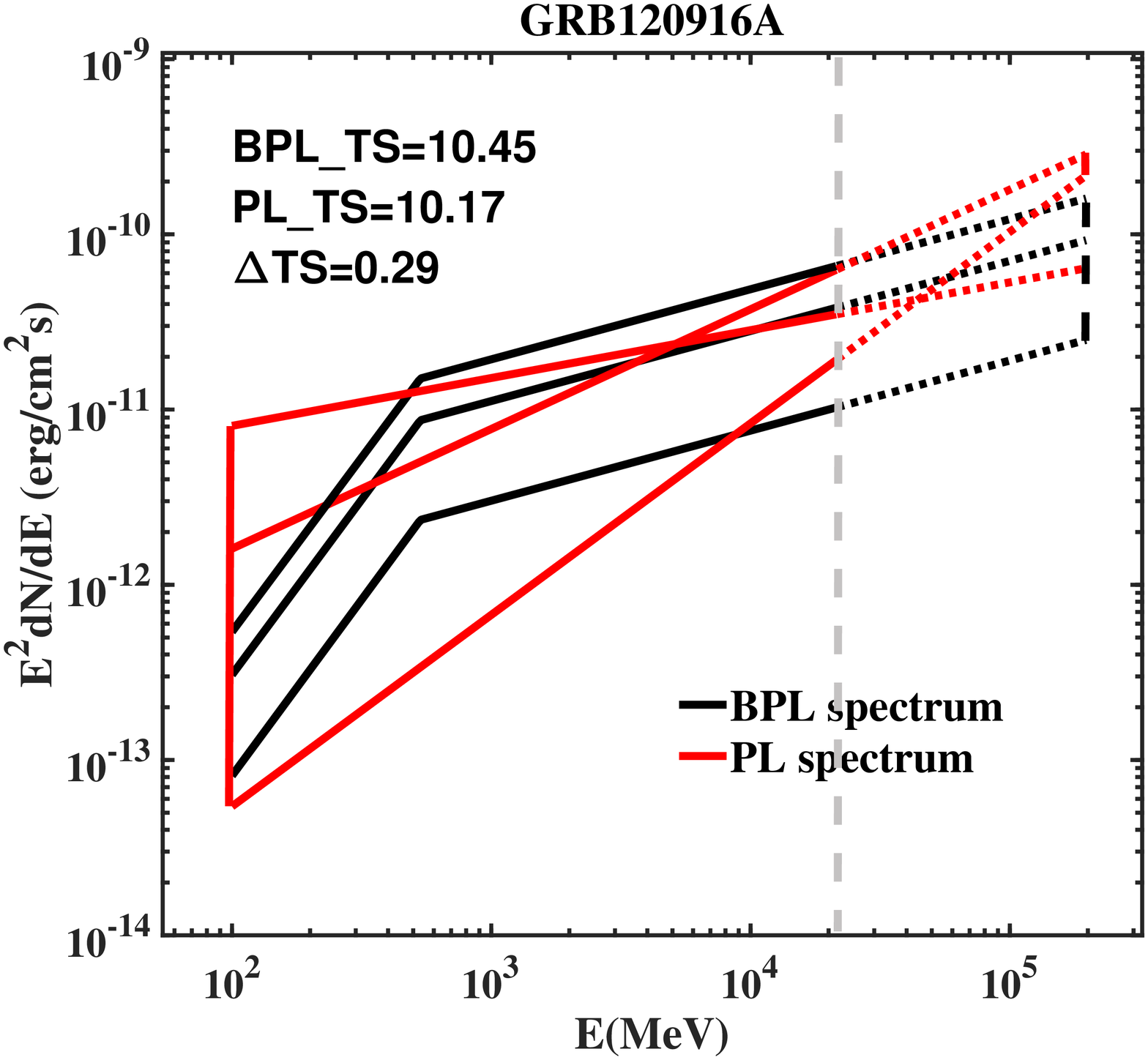}
\includegraphics[width=60mm,height=48mm]{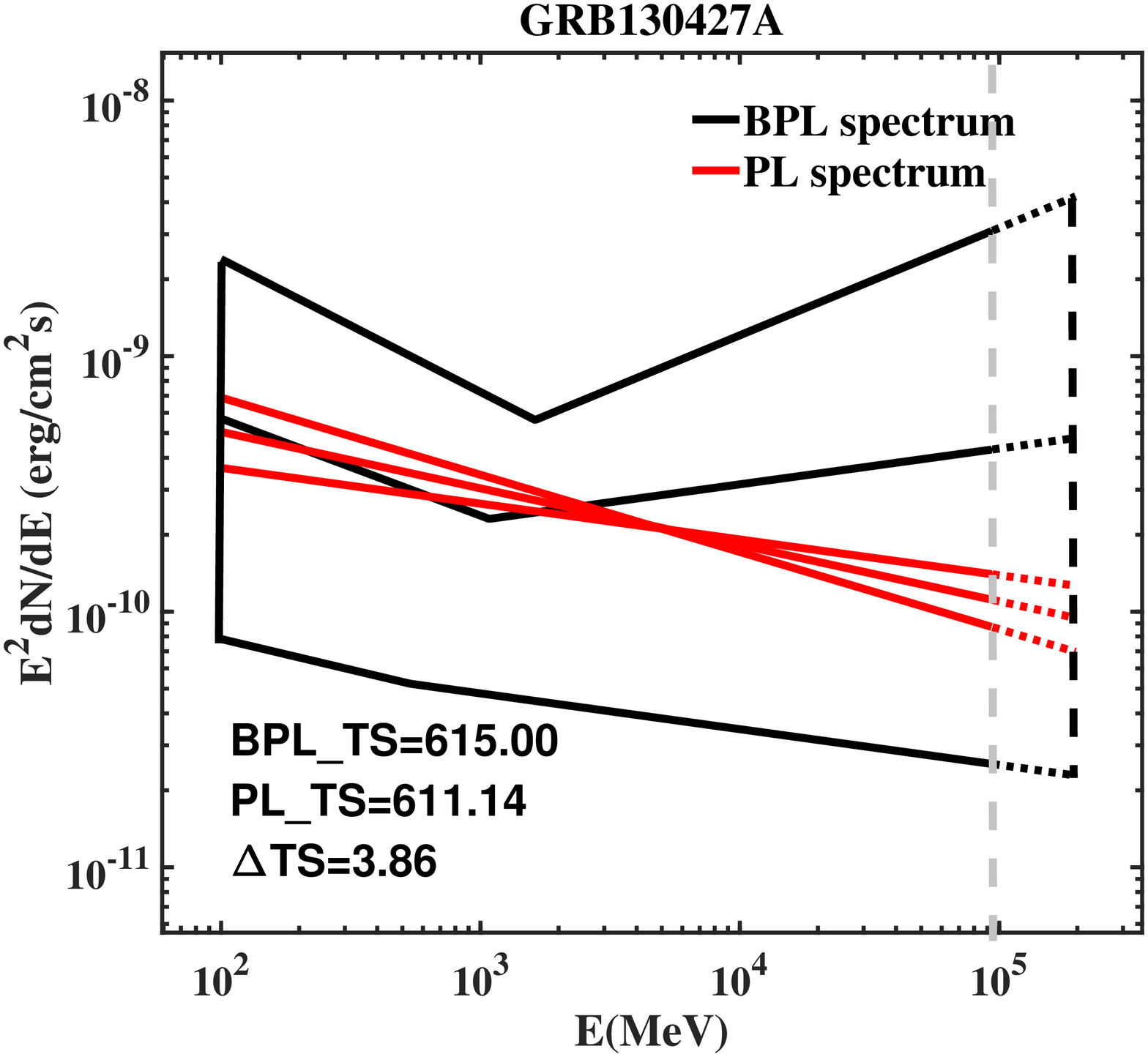}
\includegraphics[width=60mm,height=48mm]{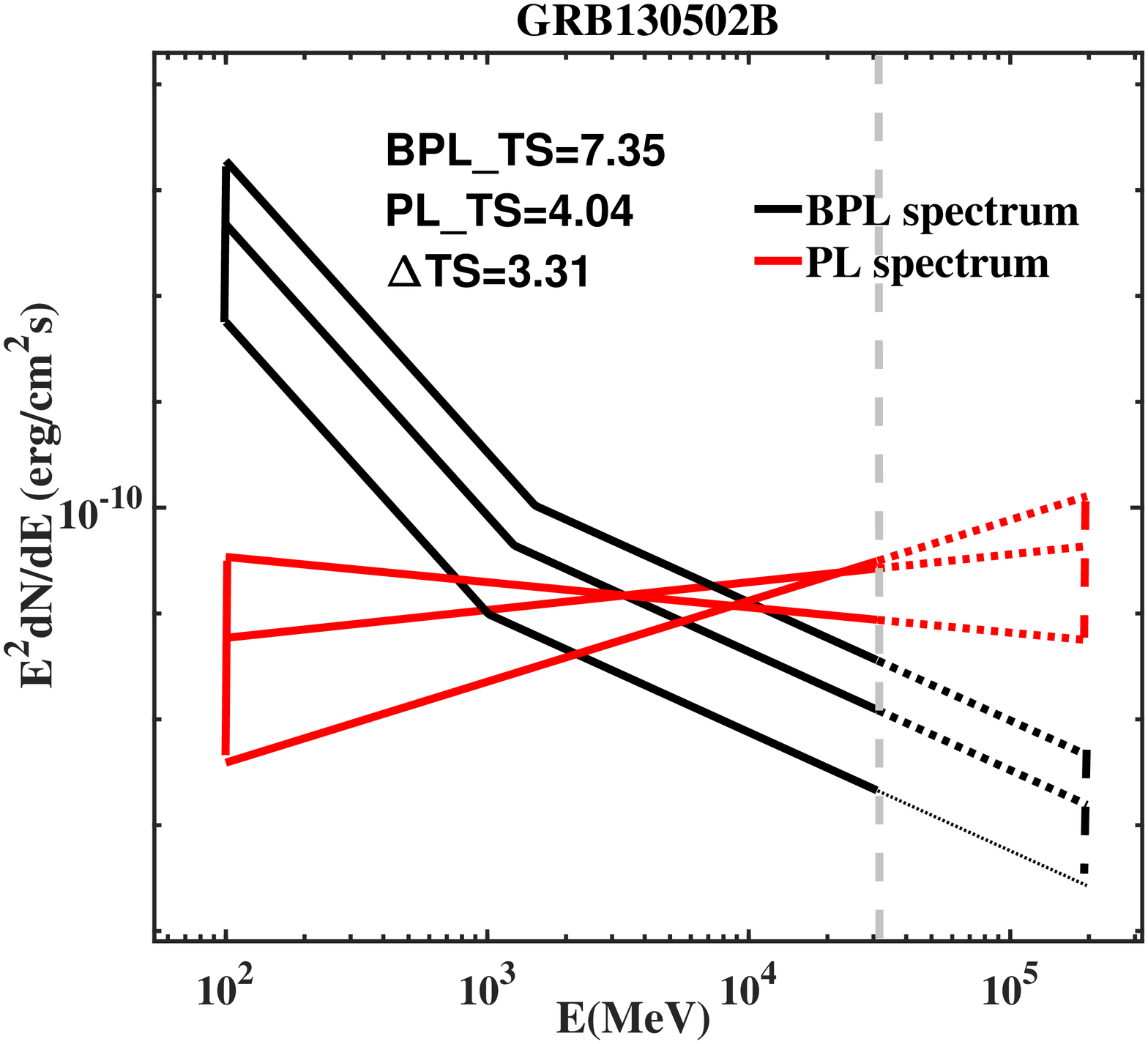}
\includegraphics[width=60mm,height=48mm]{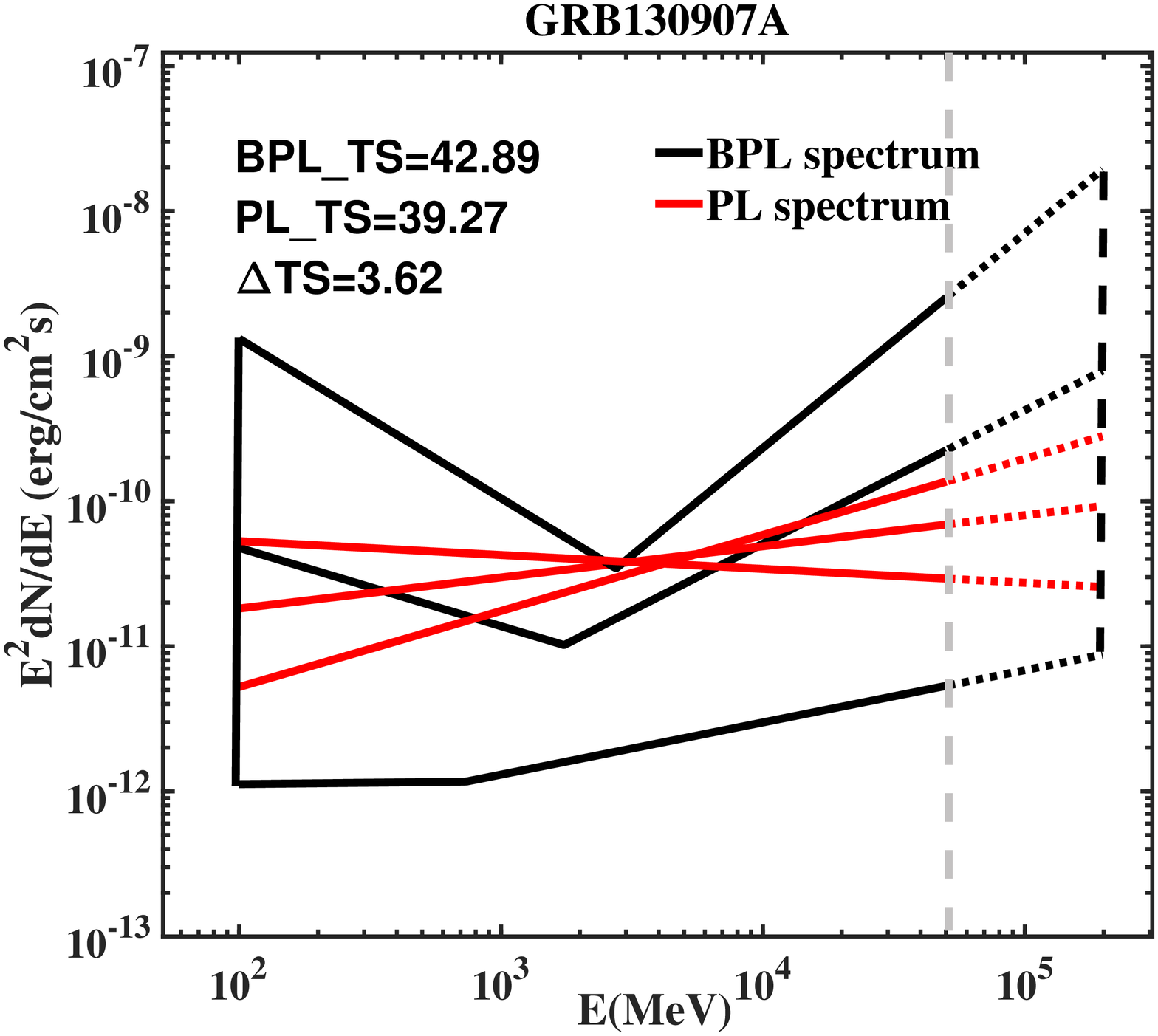}
\includegraphics[width=60mm,height=48mm]{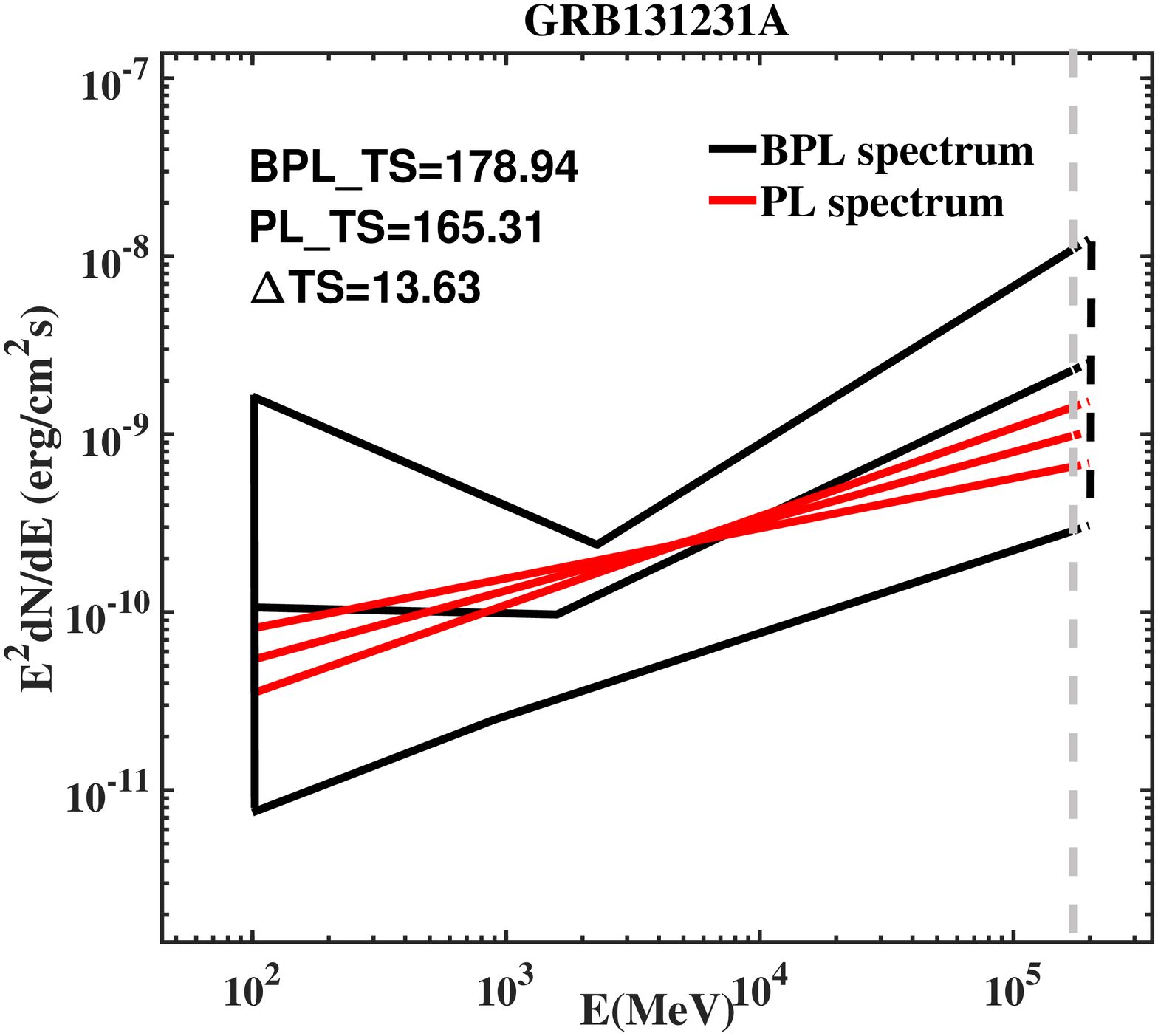}
\includegraphics[width=60mm,height=48mm]{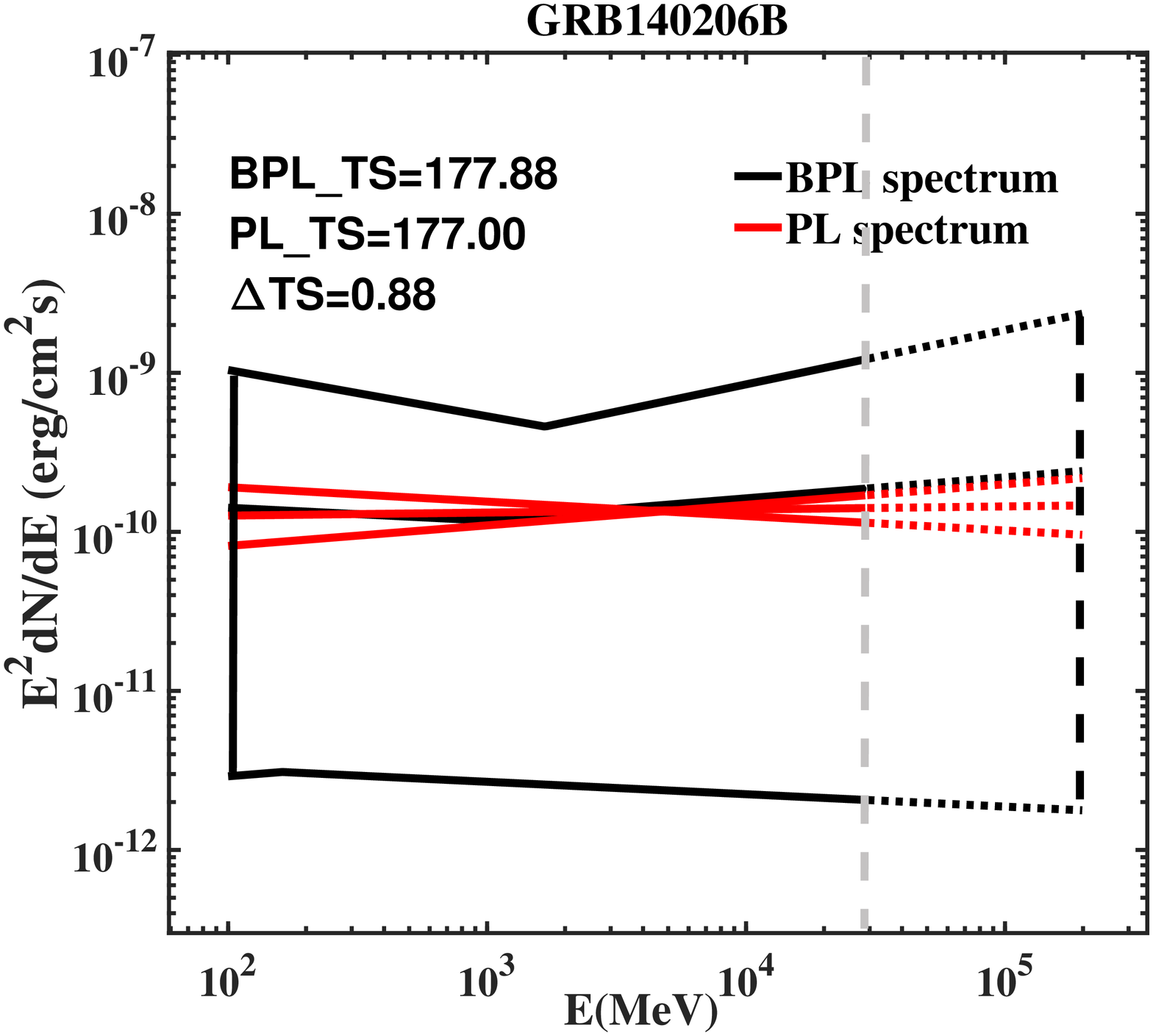}
\includegraphics[width=60mm,height=48mm]{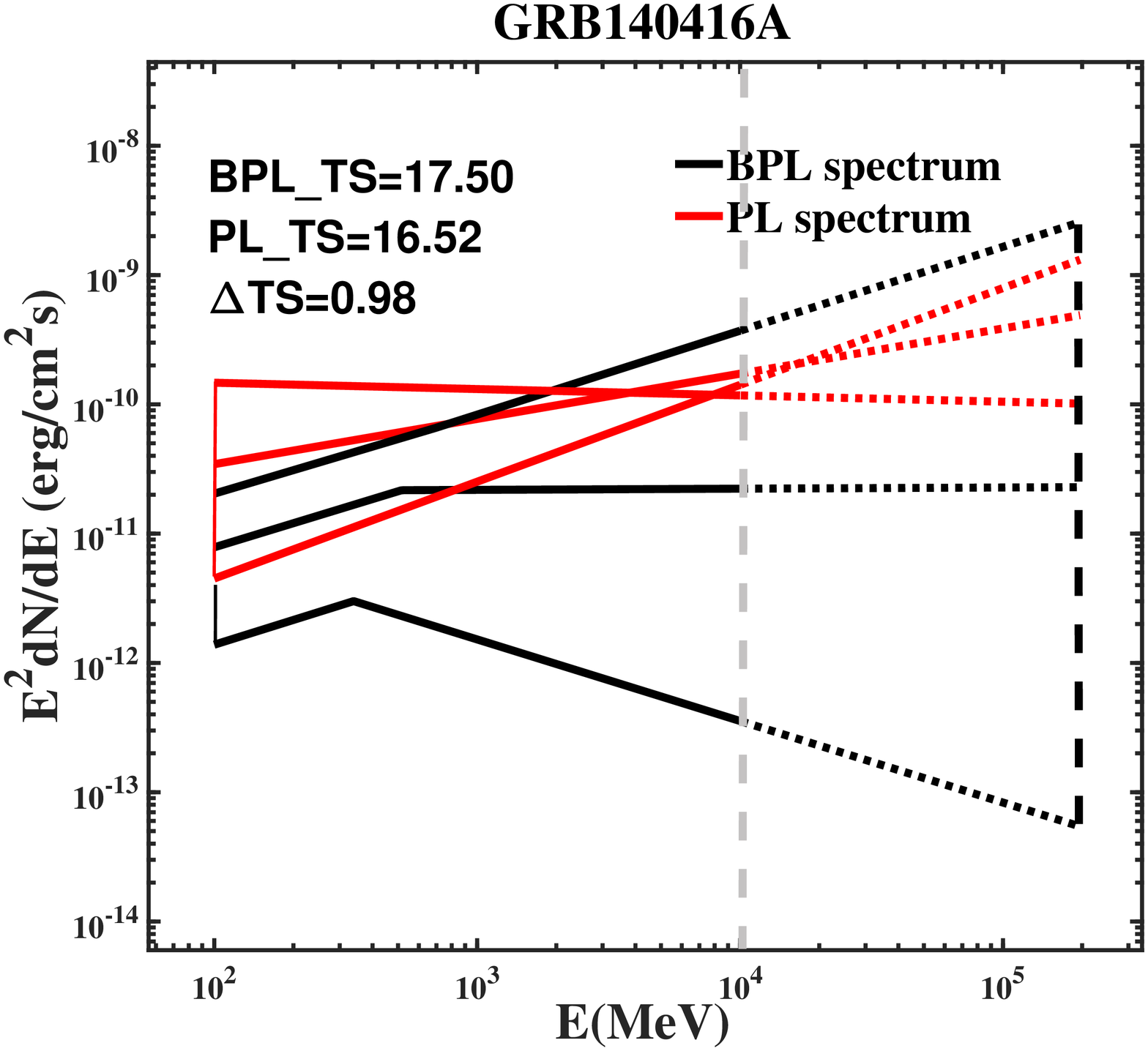}
\includegraphics[width=60mm,height=48mm]{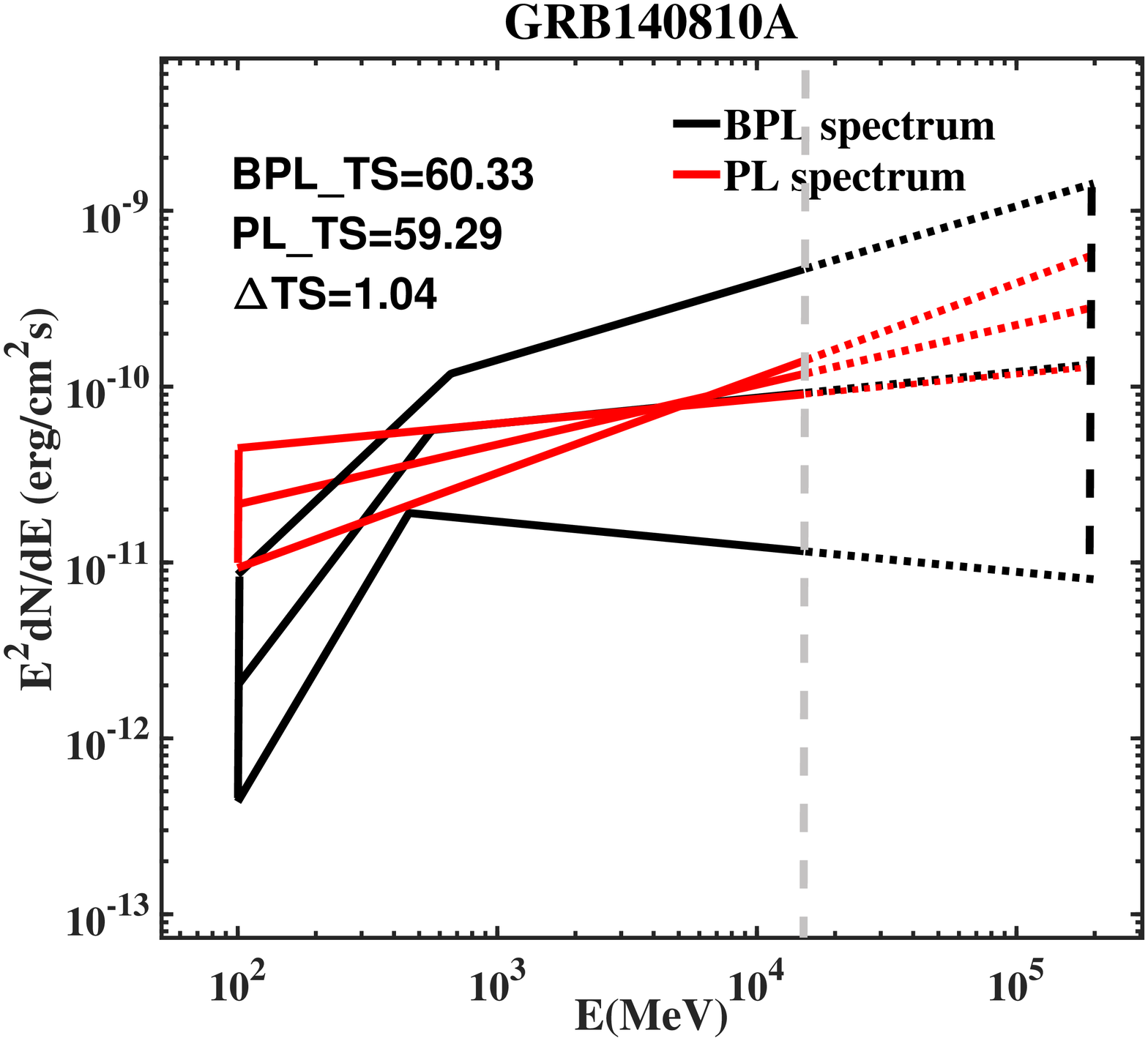}
\includegraphics[width=60mm,height=48mm]{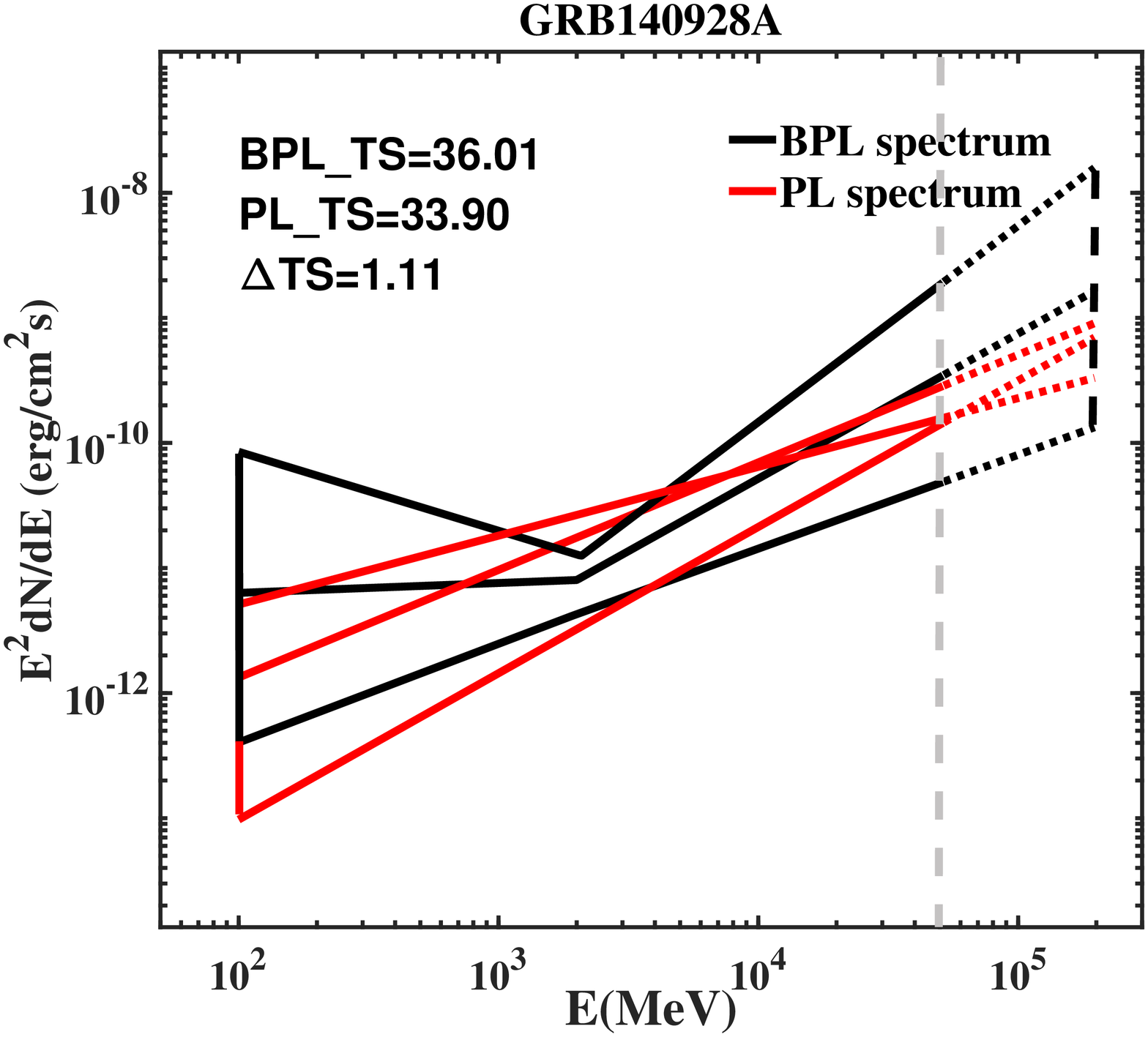}
\caption{continued in next figure}
\label{sample1}
\end{figure*}
\begin{figure*}
\includegraphics[width=60mm,height=48mm]{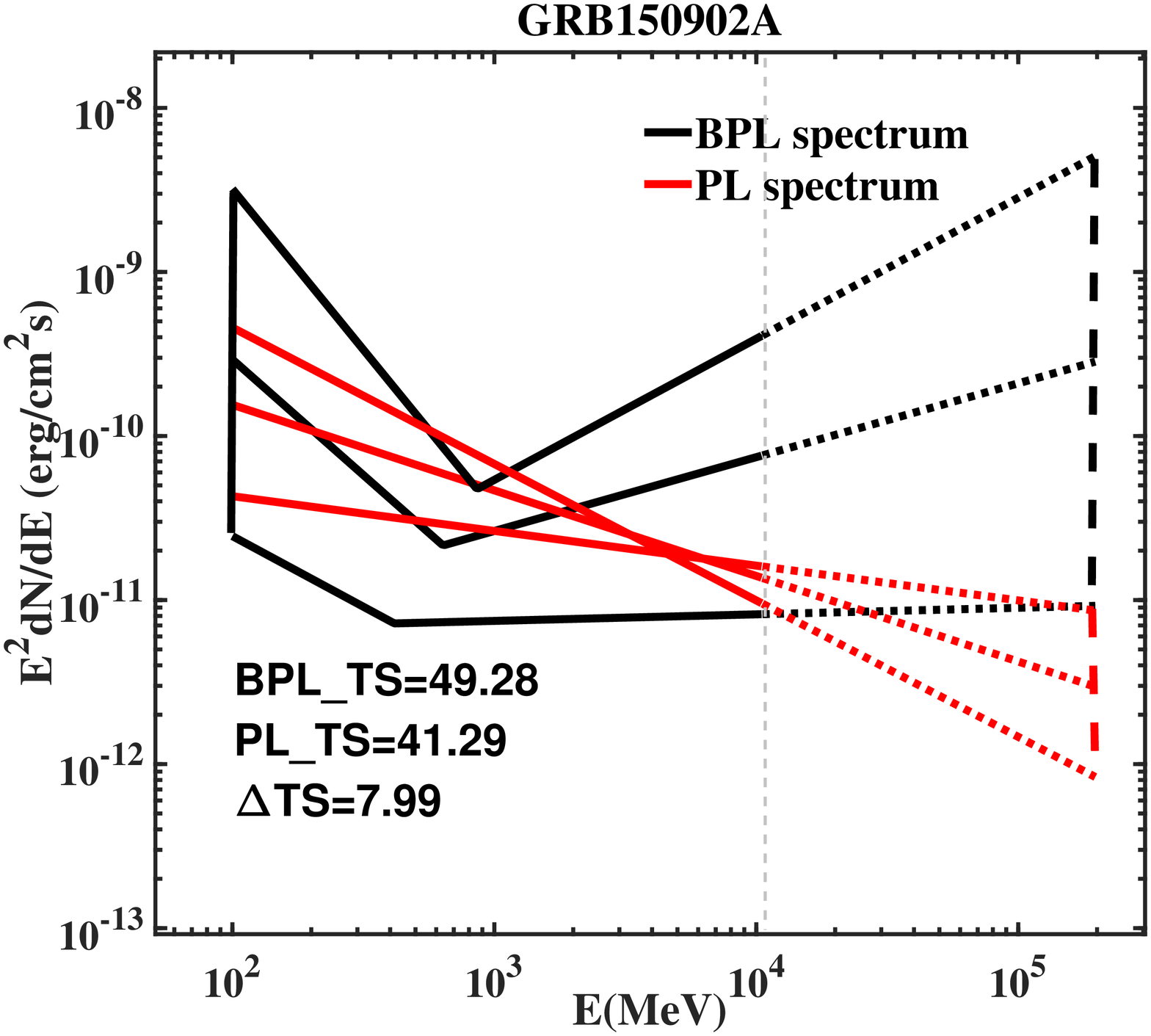}
\includegraphics[width=60mm,height=48mm]{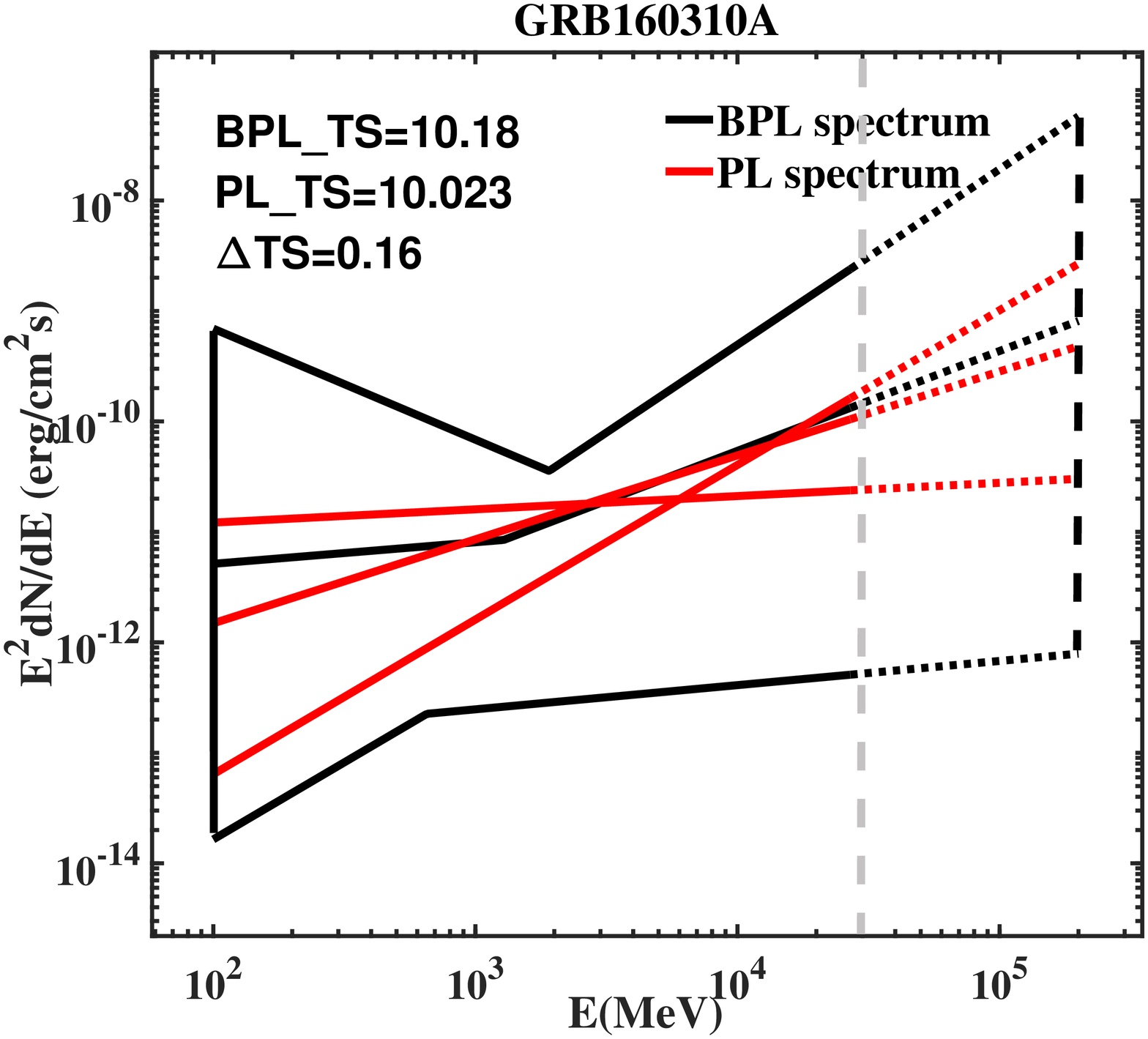}
\includegraphics[width=60mm,height=48mm]{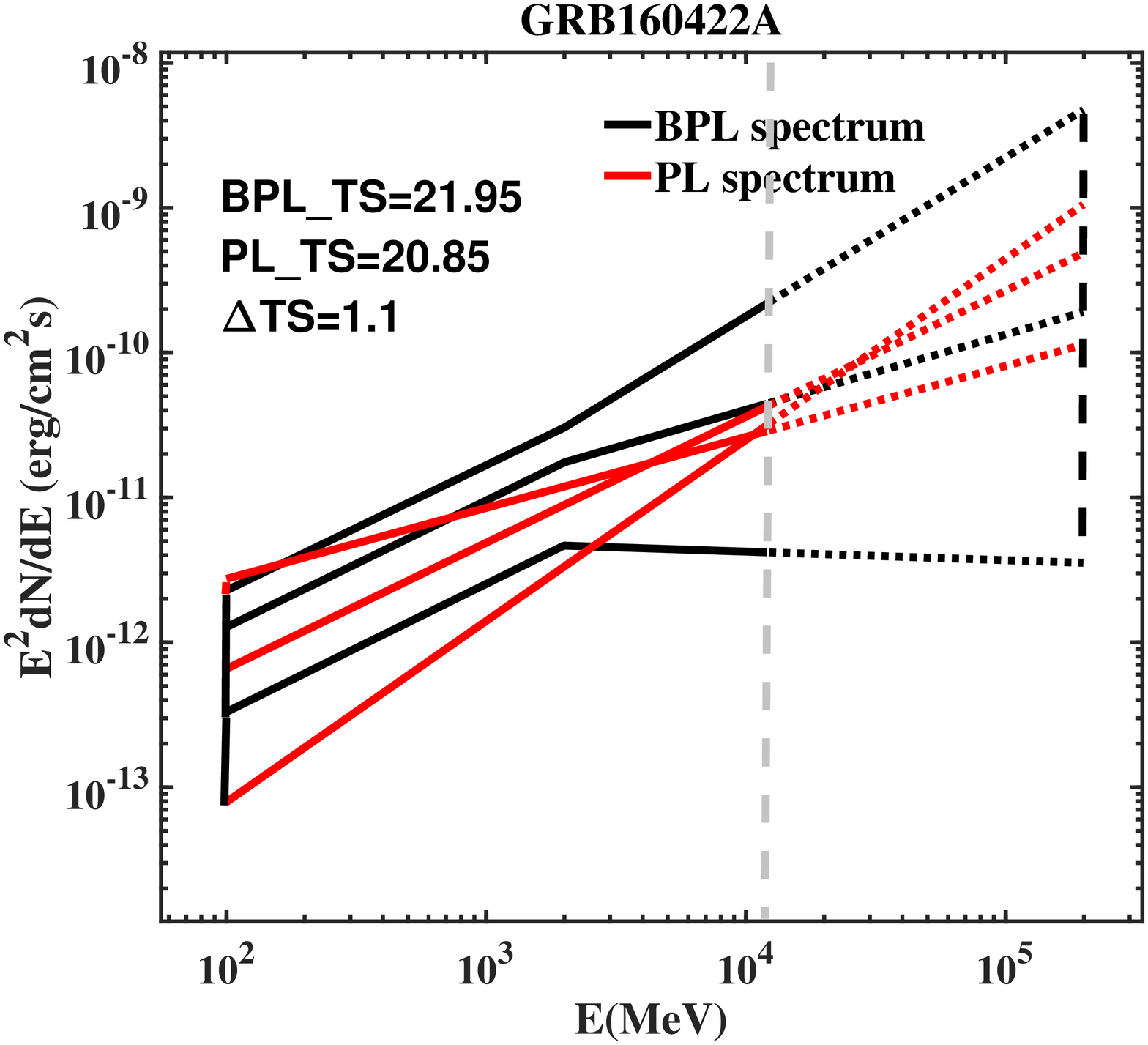}
\includegraphics[width=60mm,height=48mm]{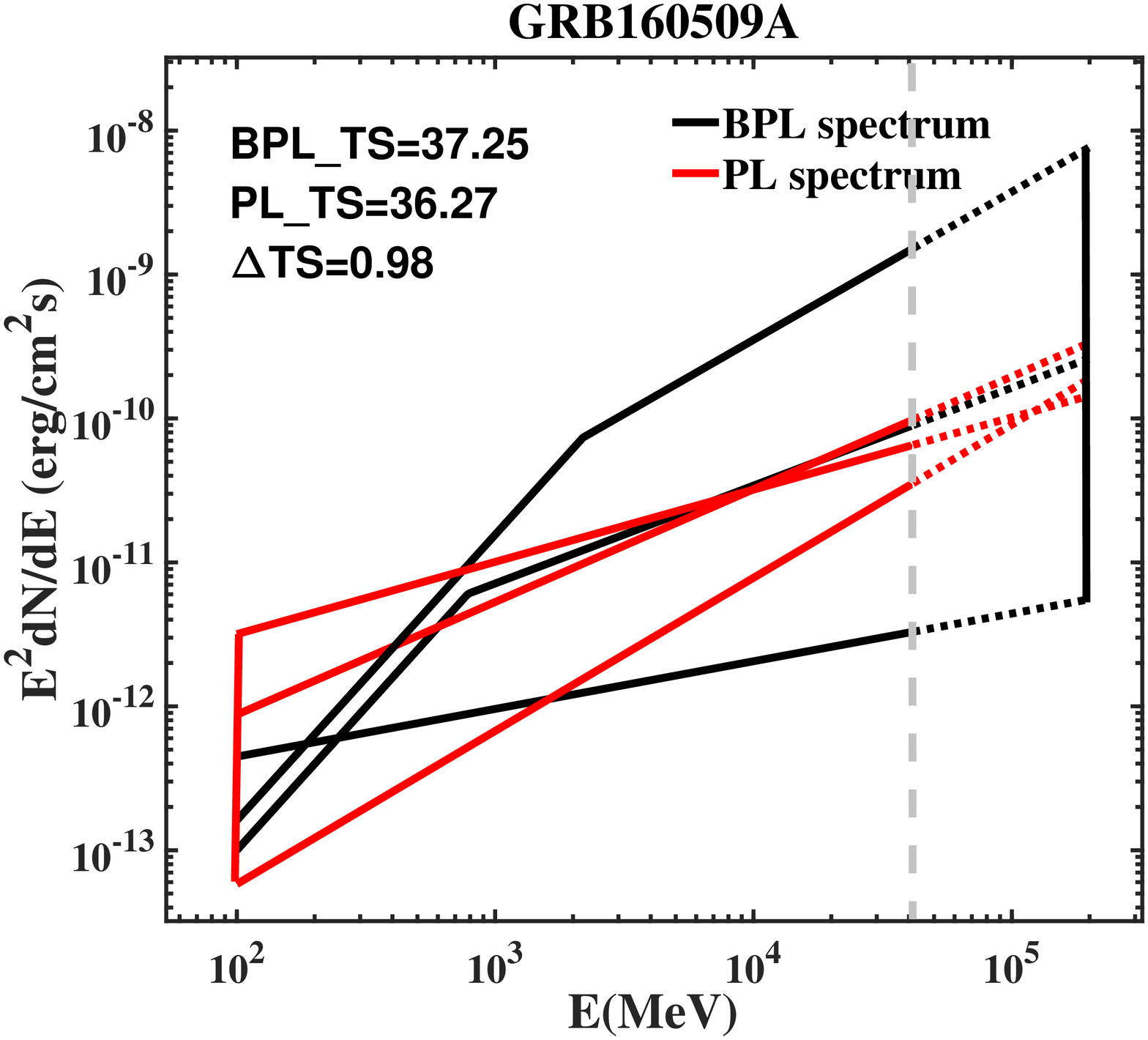}
\includegraphics[width=60mm,height=48mm]{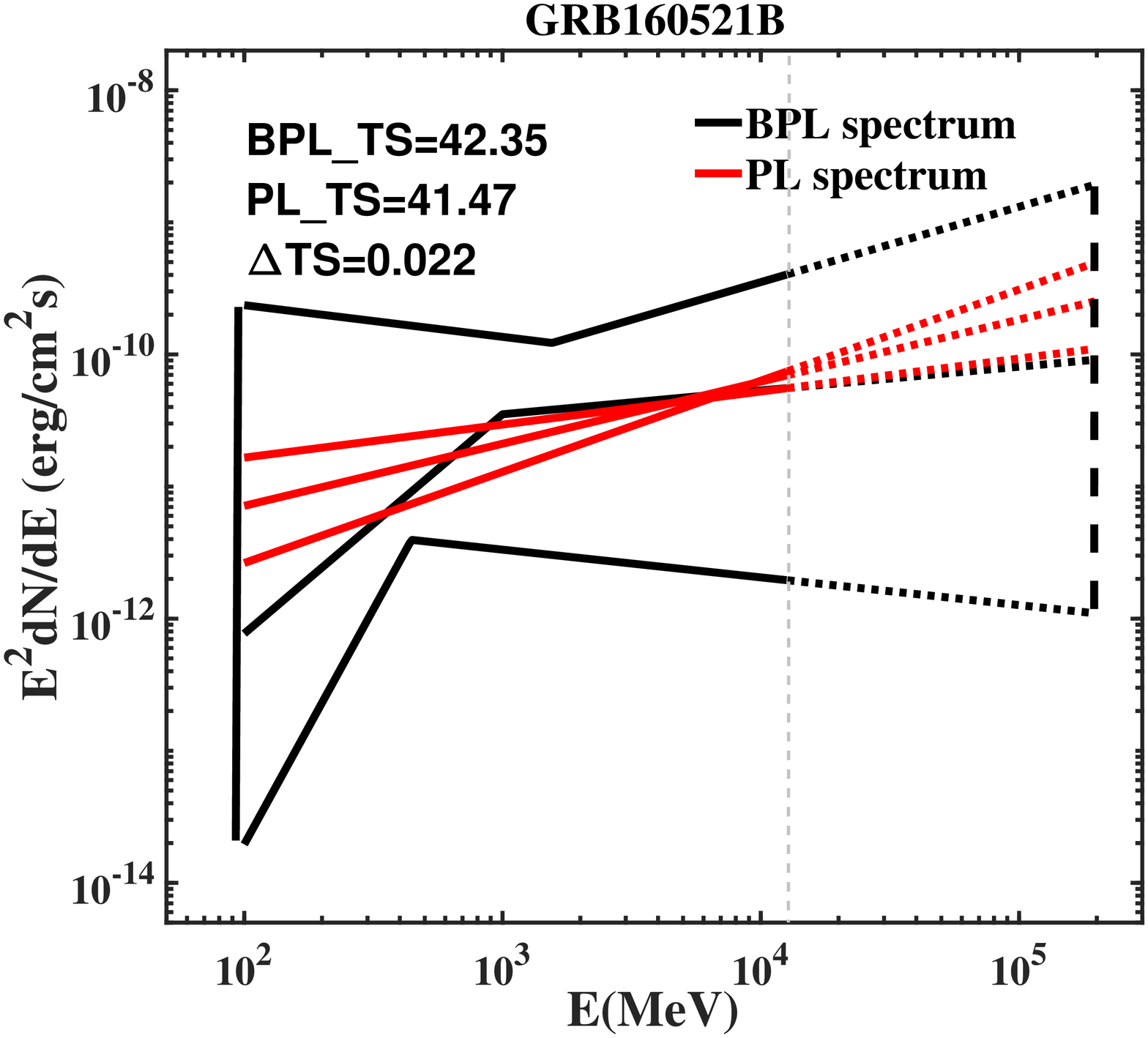}
\includegraphics[width=60mm,height=48mm]{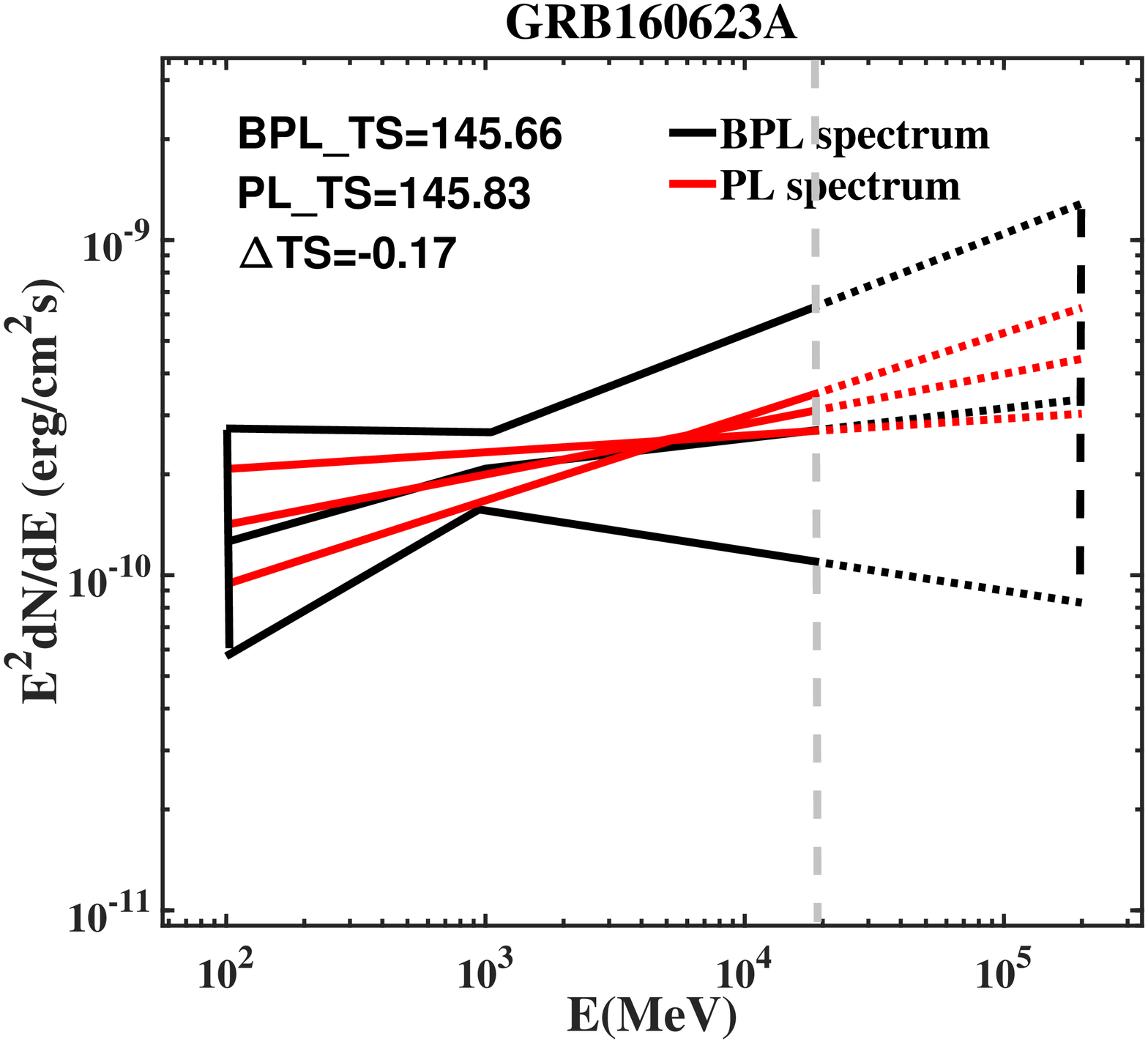}
\includegraphics[width=60mm,height=48mm]{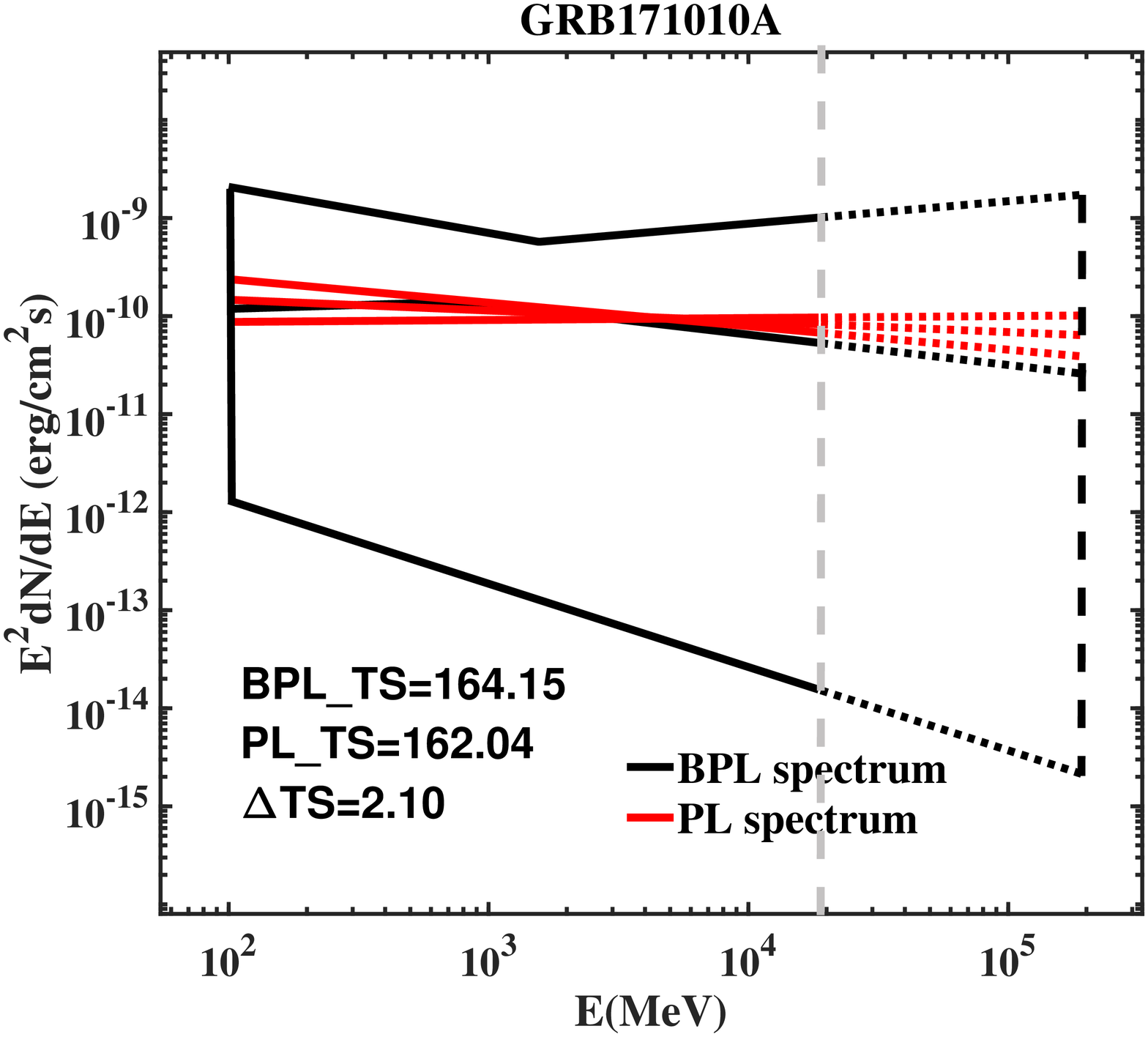}
\includegraphics[width=60mm,height=48mm]{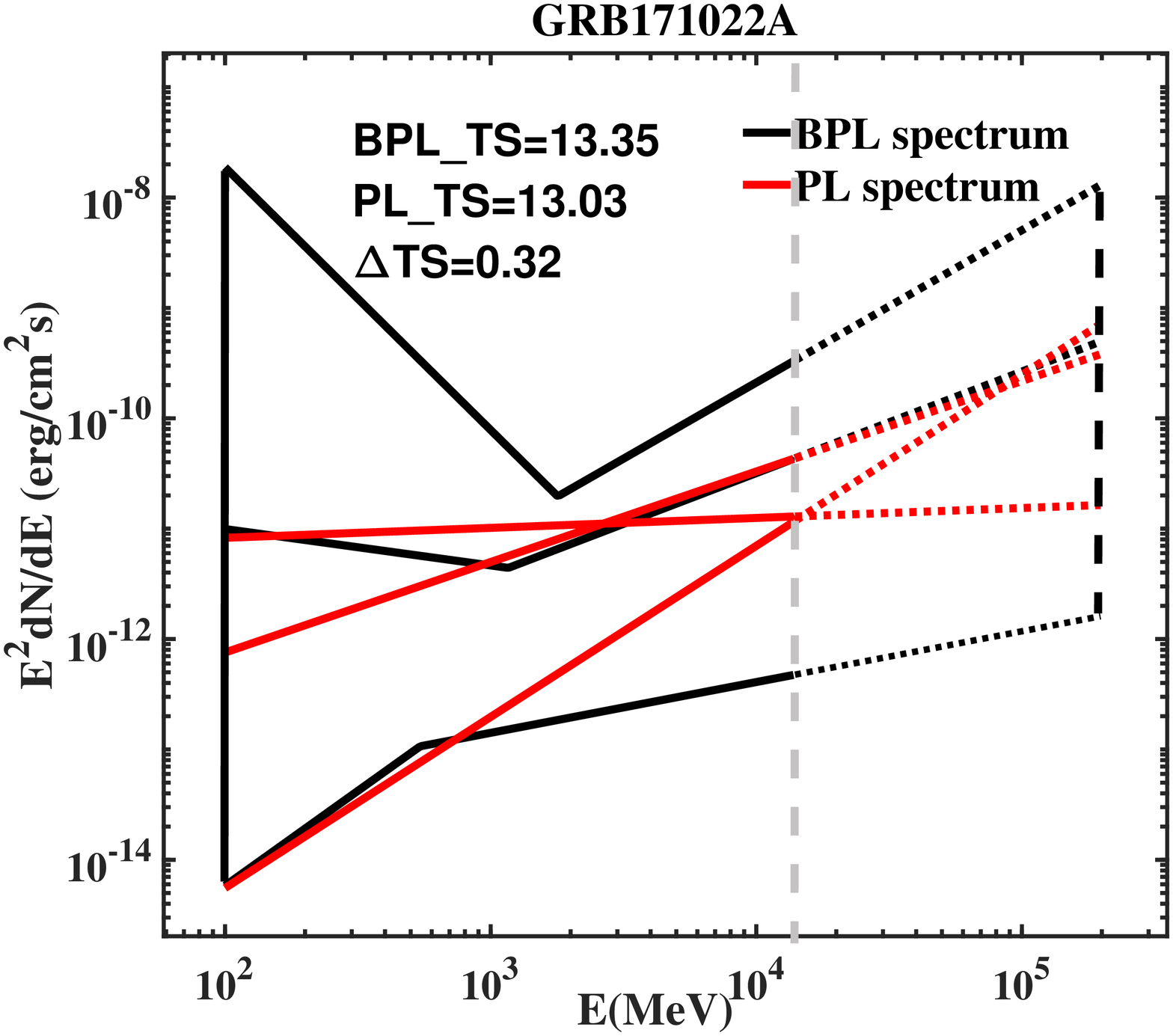}
\includegraphics[width=60mm,height=48mm]{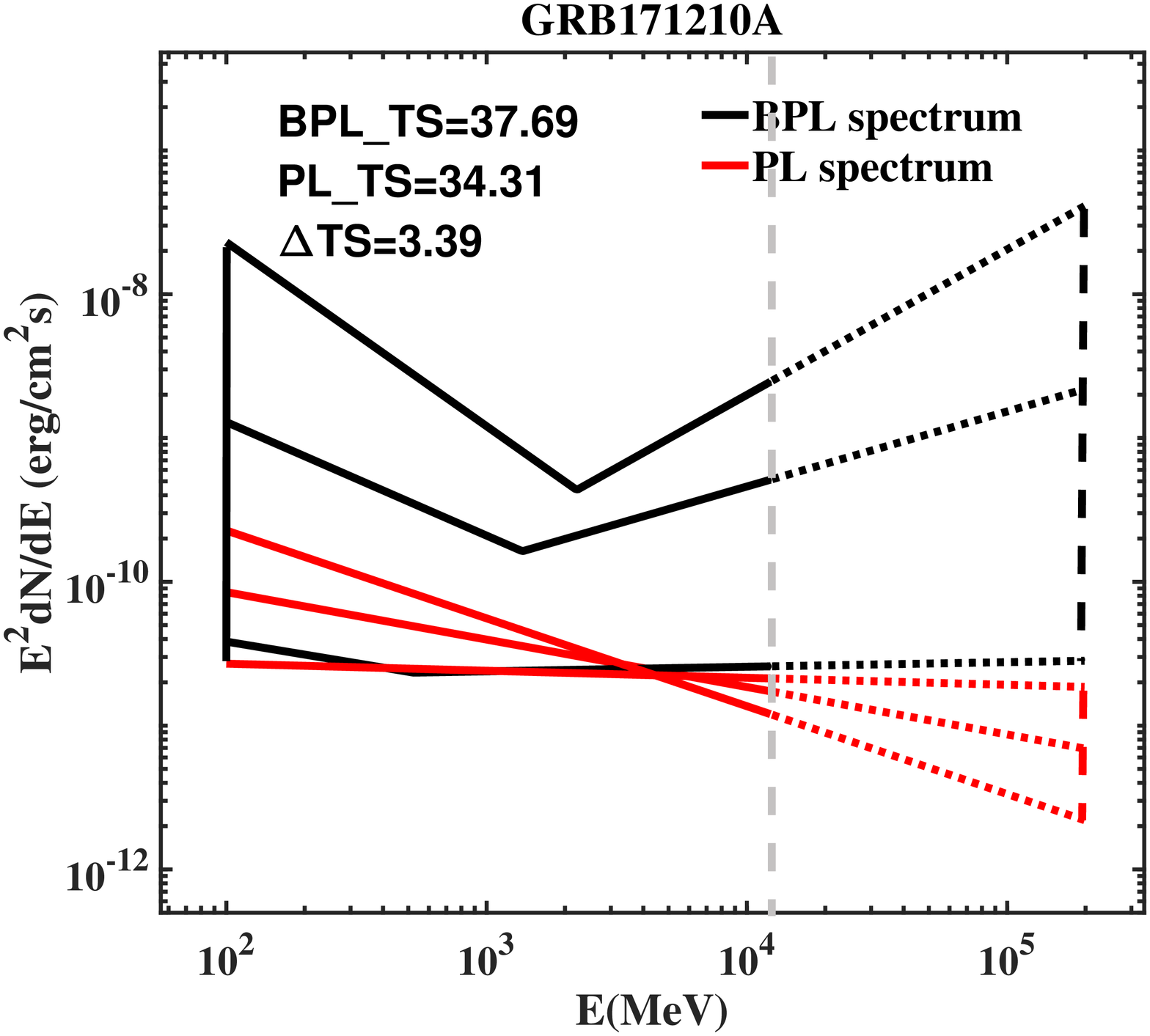}
\includegraphics[width=60mm,height=48mm]{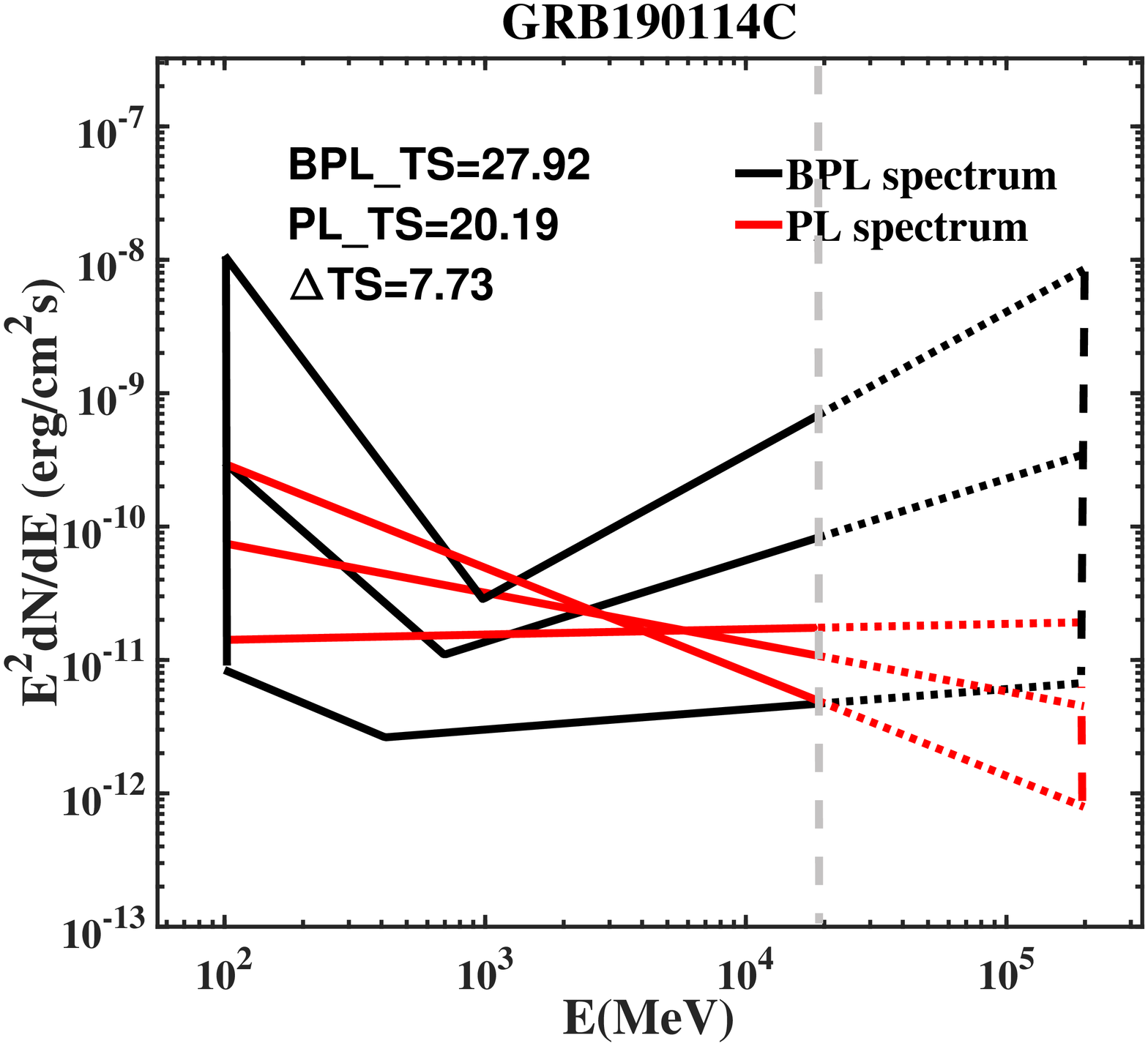}

\caption{The 0.1-200GeV spectra of the 25 ``GeV afterglows detected at $\geq$10~GeV" in the $E^2$dN/dE representation, using time range from the GRB onset to one day thereafter. Spectra are fitted with PL and BPL models. Black lines show the PL model fits, red lines show the BPL model fits. All the $\pm 1\sigma$ error contours are propagated from errors on the fit parameters. The grey dotted lines indicate the energy of the most energetic photon detected, the spectra are plotted with dotted line from the maximum photon energy to 200GeV.}
\label{sample2}
\end{figure*}

\startlongtable
\begin{deluxetable}{@{\extracolsep{4pt}}ccccccccc}
\tabletypesize{\footnotesize}
\tablecolumns{9}
\centering
\tablewidth{0pt}
\tablecaption{Highest energy events of \textit{Fermi}-LAT GRBs\label{tab_energymax_GRB}}
\tablehead{
\multicolumn{4}{c}{GRB Properties}& \multicolumn{5}{c}{Photon Properties}  \\ \cline{1-4}\cline{5-9}
%\colhead{} & \colhead{} & \colhead{} & \multicolumn{3}{c}{Observation Frame}& \multicolumn{3}{c}{Intrinsic Frame} & \colhead{} \\ \cline{4-6}\cline{7-9}
\colhead{Name} & \colhead{T90$^{a}$$_{of}$} & \colhead{T90$^{b}$$_{sf}$} &  \colhead{Redshift$^{c}$}&\colhead{Energy$^{d}$$_{of}$}  & \colhead{Arrival time$^{e}$$_{of}$} & \colhead{Energy$^{f}$$_{sf}$} & \colhead{Arrival Time$^{g}$$_{sf}$} & \colhead{P$^{h}$}}
\startdata
190114C$^{*}$ & 116.0 & 81.4 & 0.43 &18.94& 8849.2 & 27.0 & 6210.0 &1 \\
171210A$^{*}$ & 143.1 & $-$ & $-$  &12.49& 1374.6& $-$ & $-$ & 1 \\
171022A$^{*}$ &13.3 & $-$ & $-$ &13.98& 4582.8 & $-$ & $-$ & 1\\
171010A$^{*}$ & 107.3 & 80.7& 0.33&19.00& 2891.0 & 25.3 & 2173.7 & 1 \\
160623A$^{*}$& 107.8 & 78.9 &0.37 & 18.21 & 12039.8  &24.9& 8807.5 & 1 \\
160521B$^{*}$ & 2.8 & $-$ & $-$ &12.69&423.7  &$-$ & $-$ &1 \\
160509A$^{*}$ & 369.7 & 170.4   & 1.17  &51.90 & 55.7 & 112.6  & 25.7  & 1 \\
 & $-$  & $-$ &$-$ &41.51&221.4 & 90.1 & 102.0  & 1 \\
 & $-$  & $-$ & $-$ &28.90&69688.6 & 62.7 & 32114.6 & 1   \\
 160422A$^{*}$& 12.3 & $-$ & $-$ &12.29 &771.4 & $-$ & $-$ & 1  \\
160310A$^{*}$& 25.6  & $-$ & $-$ &26.99 &5886.0 & $-$ & $-$  & 1  \\
150902A$^{*}$& 13.6 & $-$ & $-$ &10.59 &98.9 & $-$ & $-$ & 1 \\
140928A$^{*}$& 17.9& $-$  & $-$ &51.76&2555.2 & $-$ & $-$ & 1 \\
 & $-$  & $-$ & $-$ &38.62 &3100.9 &$-$&$-$ & 1 \\
140810A$^{*}$&81.7 & $-$ & $-$ &15.38 &1490.3 & $-$ & $-$& 1  \\
140416A$^{*}$&31.7 & $-$ & $-$ &10.08 &2208.3& $-$ & $-$ & 1  \\
140206B$^{*}$&27.3  & $-$ & $-$ &10.96  &6736.7 & $-$ & $-$& 1 \\
 & $-$  & $-$ &$-$ &29.39 &75494.0&$-$&$-$ & 1   \\
131231A$^{*}$&31.2 &19.0 &0.64 &48.29 &110.4 & 79.3 &67.2 & 1 \\
 & $-$  & $-$  & $-$ &17.14 &844.0 & 28.1 &514.0 & 1\\
 & $-$  & $-$ & $-$&173.17 &57071.0 &284.4 &34757.0  &0.97  \\
130907A$^{*}$&115.0 &51.4  &1.24 &50.96 &17161.0 &114.1 &7668.0 & 1\\
130502B$^{*}$&24.3 & $-$ & $-$ &31.10&222.1 & $-$ & $-$ & 1 \\
130427A$^{*}$&138.2  &103.1  &0.34 &94.12 &243.6 &126.1 &181.8& 1\\
 & $-$  & $-$ & $-$ &77.11 &19.1 &103.3 &14.2 & 1  \\
 & $-$  & $-$ & $-$ &57.42 &256.7 &76.9 &191.6 & 1  \\
 & $-$  & $-$ & $-$ &38.67 &78.8 &51.8 &58.8 & 1  \\
 & $-$  & $-$ & $-$ &38.19 &3410.3 &51.2 &2545.2 & 1 \\
 & $-$  & $-$ & $-$ &33.65 &34366.6 &45.1 &25648.6 & 1  \\
 & $-$  & $-$ & $-$ &28.41 &48.0 &38.1 &35.8 & 1  \\
& $-$  & $-$ & $-$  &26.90 &85.2 &36.1 &63.6 & 1  \\
 & $-$  & $-$ & $-$ &25.36 &141.5 &34.0 &105.6 & 1  \\
 & $-$  & $-$ & $-$ &19.27 &6063.0 &25.8 &4524.9 & 1  \\
 & $-$  & $-$ & $-$ &17.08 &217.9 &22.9 &162.6 & 1  \\
 & $-$  & $-$ & $-$ &14.90 &119.7&20.0 &89.4 & 1  \\
 & $-$  & $-$ & $-$ &12.87 &80.9 &17.2 &60.4 & 1  \\
& $-$  & $-$ & $-$  &12.18 &64.9 &16.3 &48.4 & 1  \\
 & $-$  & $-$ & $-$ &12.00 &23.9 &16.1 &17.8 & 1  \\
& $-$  & $-$ & $-$  &11.74 &214.4 &15.7 &160.0 & 1  \\
 & $-$  & $-$ & $-$ &10.85 &23.7 &14.5 &17.6 & 1  \\
120916A$^{*}$&53.0 & $-$ & $-$ &21.57 &13004.8 & $-$ & $-$ & 1 \\
101014A$^{*}$&449.4  & $-$ & $-$ &13.60 &2750.7& $-$ & $-$ & 1 \\
 &$-$& $-$ & $-$ &11.20 &2962.0 & $-$ & $-$ & 1 \\
100414A$^{*}$&26.5 &11.1 &1.37 &29.80 &34.4 &70.6&14.5 & 1\\
 &$-$& $-$ & $-$ &25.13 &359.5 &59.5 &151.8 &1 \\
100213C$^{*}$&60.0  & $-$ & $-$ &34.00 &3389.0 & $-$ & $-$ & 1 \\
090926A$^{*}$&20.0 &6.4  &2.11 &19.46 &25.8 &60.5 &8.3 & 1 \\
 &$-$& $-$ &$-$&10.42 &3786.0 &32.4 &1217.4 & 1 \\
090902B$^{*}$&19.3 &6.8  &1.82 &21.72 &332.2 &61.3 &117.7 & 1\\
&$-$& $-$ & $-$  &18.11 &26.5 &51.1 &9.4 & 1 \\
&$-$& $-$ & $-$  &15.40 &45.9 &43.5 &16.3 & 1 \\
&$-$& $-$ & $-$   &14.22 &14.5 &40.1 &5.1 & 1 \\
 &$-$& $-$ & $-$  &12.66 &42.7 &35.7 &15.1& 1 \\
&$-$& $-$ & $-$   &11.89 &12.0 &33.6 &4.2& 1 \\
090510A$^{*}$&1.0 &0.5  &0.90&29.91 &1.8 &56.8 &0.9 & 1\\
160625B & 453.4 & 188.4  & 1.41&15.30& 347.5 & 36.8  &144.4 & 1 \\
140619B&2.8& $-$ & $-$ &22.74&1.1 & $-$ & $-$ & 1 \\
120919B&118.0 & $-$ & $-$ &12.70 &605.0 & $-$ & $-$ & 1 \\
120526A&43.6  & $-$ & $-$ &14.30 &1354.3 & $-$ &$-$ & 1\\
110903A&341.3 & $-$ & $-$ &15.60 &301.0 & $-$ & $-$ & 1 \\
100116A&102.5 & $-$ & $-$ &32.64 &379.2 & $-$ & $-$ & 1 \\
 &$-$& $-$ & $-$ &13.3 &296.7 & $-$ & $-$ & 1 \\
100511A &42.4 & $-$ & $-$ &46.00 &161.9& $-$ & $-$ & 1\\
 &$-$& $-$ & $-$ &18.00 &179.8 & $-$ & $-$ & 1 \\
090427A&12.3 & $-$ & $-$ &14.10 &422.9 & $-$ & $-$ & 1 \\
080916C&63.0 &11.8  &4.35 &12.42 &17.2 &66.5 &3.2 & 1\\
&$-$& $-$ &$-$&27.43 &41.1&146.7 &7.7  & 1 \\
 \enddata
\tablenotetext{}{  \,$^{*}$: GeV afterglow GRBs}
\tablenotetext{}{ \,the unit of  Energy is GeV, the units of  T90 and Arrive time are s}
\tablenotetext{}{ \,a,d,e: Observation Frame($_{of}$), b,f,g: Source Frame($_{sf}$)}
\tablenotetext{}{ \,a$^{ref}$: https://heasarc.gsfc.nasa.gov/cgi-bin/W3Browse/w3hdprods.pl }
\tablenotetext{}{ \,c$^{ref}$: http://www.mpe.mpg.de/~jcg/grbgen.html }
\tablenotetext{}{ \,$^{h}$P: probability of the photon being associated with that GRB}
\end{deluxetable}


\begin{thebibliography}{}
\bibitem[Abdalla \etal(2019)]{HESS} Abdalla, H., Adam, R., Aharonian, F.~A. \etal, 2019, Nature, 575, 466\
\bibitem[Abdo \etal(2009)]{Abdo2009} Abdo, A., Ackermann, M., Arimoto, M. \etal, 2009b, Sci, 323, 1688\
\bibitem[Ackermann \etal(2012)]{Fermi} Ackermann, M. et al., 2012, ApJS, 203, 4\
\bibitem[Ackermann \etal(2013)]{lat_grb_cat} Ackermann, M. \etal (Fermi/LAT collaboration) 2013, \apjs, 209, 11\
\bibitem[Ackermann \etal(2014)]{130427A} Ackermann, M., Ajello, M., Asano, K. \etal, 2014, Science , 343, 6166\
\bibitem[Ajello \etal(2019)]{lat_grb_cat2} Ajello, M. Arimoto, M., Axelsson, M. \etal, 2019, \apj , 878, 52\
\bibitem[Aliu \etal(2014)]{aliu14} Aliu, E., Aune, T., Barnacka, A.\etal, 2014, \apjl, 795, L3\ 
\bibitem[Atwood \etal(2009)]{Fermi-LAT} Atwood, W.~B., et al., 2009, \apj , 697, 1071\
\bibitem[Barniol Duran and Kumar(2011)]{Duran2011} Barniol Duran, R., \& Kumar. P., 2011, MNRAS., 412, 522\
\bibitem[Bai \etal(2019)]{lhaaso} Bai, X., Bi, B. Y., Bi, X. J. \etal\ (The LHAASO collaboration), 2019, preprint[arXiv:1905.02773]
\bibitem[The Cherenkov Telescope Array Consortium(2018)]{CTA} The Cherenkov Telescope Array Consortium, 2018, preprint[arXiv:1709.07997]
\bibitem[Costa \etal(1997)]{Costa97} Costa, E. \etal 1997a, IAU Circular No. 6572\
\bibitem[Derishev and Piran(2019)]{Derishev2019} Derishev, E., \& Piran, T., arXiv:1905.08285\
\bibitem[Fan and Piran(2008)]{Fan2008} Fan, Y. Z., \& Piran, T., 2008, Fron. Phys. China., 3, 306\
%\bibitem[Fraija \etal(2019)]{Fraija2019} Fraija, N., Dichiara, S., Caligula. A. C. \etal , 	arXiv:1905.13572\
\bibitem[Galli and Piro(2008)]{Galli2008} Galli. A., \& Piro, L., 2008, A\&A., 489, 1073\
\bibitem[GCN Circular(2019)]{GCN25566} https://gcn.gsfc.nasa.gov/gcn3/25566.gcn3
\bibitem[Klebesadel and Strong\etal(1973)]{Klebesadel73} Klebesadel, R. W., Strong, I. B., \& Olson, R. A.\ 1973, \apj, 182, L85\
\bibitem[Kumar \& Barniol Duran(2010)]{KBD10} Kumar, P. \& Barniol Duran, R.\ 2010, MNRAS, 409, 226
\bibitem[Lemoine(2015)]{Lemoine2015} Lemoine, M., 2015, MNRAS, 453, 3772\
%\bibitem[Li \etal(2019)]{Li2019} Li, Liang; Geng, Jin-Jun; Meng, Yan-Zhi, 2019, \apj, 884, L109\
\bibitem[Liu \etal(2014)]{Liu2014} Liu, B., Chen, W., Liang, Y.-F.\ 2014, ApJL, 787, L6
\bibitem[MAGIC Collaboration \etal(2019a)]{MagicA} MAGIC Collaboration*, V. A. Acciari, S.Ansoldi \etal, 2019, Nature, 575, 455\
\bibitem[MAGIC Collaboration \etal(2019b)]{MagicB} MAGIC Collaboration, P. Veres, and P. N. Bhat \etal, 2019, Nature, 575, 459\
\bibitem[Meegan \etal(1998)]{Meegan1998} Meegan, C. A., Paciesas, W. S., Pendleton, G.~N., et al., 1998, AIP Conf. Proc., 428, 3\
\bibitem[Meegan \etal(2009)]{Fermi-GBM} Meegan, C.~A., et al., 2009, \apj , 702, 791\
\bibitem[M\'esz\'aros and Rees(1997)]{Meszaros97} M\'esz\'aros, P. \& Rees, M.~J., 1997, \apj , 476, 232\
\bibitem[M\'esz\'aros and Rees(2000)]{Meszaros2000} M\'esz\'aros, P. \& Rees, M.~J., 2000, \apj , 541, L5\
\bibitem[M\'esz\'aros \etal(2004)]{Meszaros2004} M\'esz\'aros, P., Razzaque, S., \& Zhang, B., 2004, New Astron. Rev., 48, 445\
\bibitem[Nakar \etal(2009)]{Nakar2009} Nakar, E., Ando, S., \& Rari, R., 2009, APJ., 703, 675\
\bibitem[Panaitescu (2017)]{Panaitescu2017} Panaitescu. A , 2017, \apj, 837, 13\
\bibitem[Panaitescu and M\'esz\'aros(1998)]{Panaitescu1998} Panaitescu. A., \& M\'esz\'aros, P., 1998, ApJ, 501, 772\
\bibitem[Piran \& Nakar(2010)]{PiranNakar10} Piran, T. \& Nakar, E., 2010, \apjl, 718, L63
\bibitem[Rees and M\'esz\'aros(1992)]{Ree92} Rees, M.~J. \& M\'esz\'aros, P.\ 1992, \mnras , 258.41P\
\bibitem[Rees and M\'esz\'aros(1994)]{Ree94} Rees, M.~J. \& M\'esz\'aros, P.\ 1994, \apj , 430.L93\
\bibitem[Sari and Esin(2001)]{Sari2001} Sari, R., \& Esin, A. A., 2001, ApJ, 548, 787\
\bibitem[Sari \etal(1998)]{Sari1998} Sari, R., Piran, T., \& Narayan, R., 1998, ApJ, 497, L17\
\bibitem[Tam \etal(2017)]{Tam2017} Tam, P. -H. T., He, X. B, Tang, Q.-W.\etal, 2017, ApJL, 844, L2\
\bibitem[Tam \etal(2013)]{Tam2013} Tam, P. -H. T., Tang, Q.-W., Hou, S.-J \etal, 2013, ApJL, 771, L13\
\bibitem[Tang \etal(2014)]{Tang2014} Tang, Q.-W., Tam, P. -H. T. \& Wang, X.Y. , 2014, ApJ, 788, 156\
\bibitem[Wang \etal(2019)]{Wang2019} Wang, X. Y., Liu, R. Y., Zhang, H. M. \etal , arXiv:1905.11312\
\bibitem[Wei and Lu(1998)]{Wei1998} Wei, D. M., \& Lu, T., 1998, ApJ, 505, 252\
\bibitem[Yassine \etal(2017)]{Yassine2017} Yassine, M., Piron, F., Mochkovitch, R. \etal , 2017, A\&A, 606, 17\
\bibitem[Zhang \etal(2019)]{Zhang2019} Zhang. B., Christie. I. M., Petropoulou, M. \etal , arXiv:1910.14049\
\bibitem[Zhang and M\'esz\'aros(2001)]{Zhang2001} Zhang, B. \& M\'esz\'aros, P.\ 2001, \apj , 559, 110\
    
%\acknowledgments
%Support by

\end{thebibliography}
\end{document}